\documentclass[11pt,dvips,twoside,letterpaper]{article} 
\usepackage{times,fancyhdr}
\usepackage{graphicx}
\usepackage{geometry}
\usepackage{graphics,amssymb}


\makeatletter 
\def\cleardoublepage{\clearpage\if@twoside \ifodd\c@page\else%
    \hbox{}%
    \thispagestyle{empty}
    \newpage%
    \if@twocolumn\hbox{}\newpage\fi\fi\fi} 
\makeatother

\begin{document}
\title{
{\begin{flushleft}
\vskip 0.45in
\end{flushleft}
\vskip 0.45in
\bfseries\scshape Dynamical Simulation of Nuclear ``Pasta'':\\
Soft Condensed Matter in Dense Stars}}
\author{\bfseries\itshape Gentaro Watanabe$^{a,b}$\footnote{
Present address: CNR-INFM BEC Center, Department of Physics, 
University of Trento, Via Sommarive 14, 38050 Povo (TN) Italy}~\ and 
Hidetaka Sonoda$^{c,b}$\\
$^{a}$NORDITA, Blegdamsvej 17, DK-2100 Copenhagen \O, Denmark\\
$^{b}$The Institute of Chemical and Physical Research (RIKEN)\\
Saitama 351-0198, Japan\\
$^{c}$Department of Physics, University of Tokyo,
Tokyo 113-0033, Japan}
\date{}
\maketitle
\thispagestyle{empty}
\setcounter{page}{1}
\thispagestyle{fancy}
\fancyhead{}
\fancyhead[L]{In: Soft Condensed Matter: New Research \\ 
Editor: Kathy I. Dillon, pp. {\thepage-\pageref{lastpage-01}}}
\fancyhead[R]{ISBN 1-59454-665-7  \\
\copyright~2006 Nova Science Publishers, Inc.}
\fancyfoot{}
\renewcommand{\headrulewidth}{0pt}

\begin{abstract}
More than twenty years ago, it was predicted 
that nuclei can adopt interesting shapes,
such as rods or slabs, etc., 
in the cores of supernovae and the crusts of neutron stars.
These non-spherical nuclei are referred to as nuclear ``pasta.''

In recent years, we have been studying the dynamics of the pasta phases using 
a method called quantum molecular dynamics (QMD)
and have opened up a new aspect of study for this system.
Our findings include: dynamical formation of the pasta phases 
by cooling down the hot uniform nuclear matter;
phase diagrams in the density versus temperature plane;
structural transitions between the pasta phases 
induced by compression and elucidation of the mechanism by which they proceed.
In the present article, we give an overview of the basic physics and astrophysics
of the pasta phases and review our works for readers in other fields.
\end{abstract}



\section{Introduction \label{SN and NS}}

Massive stars with masses $M=8- (30-50)M_\odot$ 
($M_\odot$ is the mass of the sun) end their lives in 
the most spectacular and violent events in the Universe:
explosions called supernovae, which are driven by 
gravitational collapse of the stellar core 
(see, e.g., Ref.\ \cite{shapiro}).
The mechanism for the collapse-driven supernova explosion
has been a central mystery in astrophysics for almost half a century
\cite{colgate}.

Great efforts have been made to unravel the mystery
and a qualitative picture has been obtained so far
(see, e.g., Refs.\ \cite{bethe,janka,suzuki,wilson,whw,woosley}).
We now know that after supernova explosions the collapsing core 
becomes a dense compact object, neutron star, for $M=8- 30 M_\odot$ 
or further collapses to a black hole for $M\gtrsim 30 M_\odot$.
However, no calculations incorporating reasonable physical input
succeeded in reproducing an explosion;
in the calculations, shock waves stall on their way out of the collapsing core
and the core does not explode but contracts into a black hole.
These earlier studies suggest that
interactions between neutrinos from the central region of the star
and the shocked matter
are important for the success of an explosion.
This will hopefully revive the shock wave, and lead to a successful explosion.

In equilibrium dense matter in supernova cores and neutron stars,
the existence of nuclei with rod-like and slab-like shape
is predicted \cite{hashimoto,rpw}. Such nuclei with exotic shapes
are referred to as nuclear ``pasta.''
The nuclear pasta itself is an interesting subject from the point of view
of materials science of dense stellar matter and it has been studied by
nuclear physicists in addition to astrophysicists for more than twenty years.
Furthermore, the pasta phases 
have recently begun to attract the attention of many researchers
(see, e.g., Refs.\ \cite{burrows,martinez} and references therein).
As has been pointed out in Refs.\ \cite{sato_presen,gentaro2,qmd1}
and elaborated in Refs.\ \cite{horowitz2,horowitz3,sonoda},
the existence of the pasta phases modifies the interaction
between neutrinos and matter significantly.
Our recent work \cite{qmd_transition}
strongly suggests the possibility of dynamical formation
of the pasta phases in collapsing cores from a crystalline lattice of spherical nuclei;
effects of the pasta phases on the supernova explosions
should be seriously discussed in the near future.

\pagestyle{fancy}
\fancyhead{}
\fancyhead[EC]{Gentaro Watanabe and Hidetaka Sonoda}
\fancyhead[EL,OR]{\thepage}
\fancyhead[OC]{Dynamical Simulation of Nuclear ``Pasta'': Soft Condensed Matter...}
\fancyfoot{}
\renewcommand\headrulewidth{0.5pt} 

In the present article, we provide an overview of the physics
and astrophysical background of the pasta phases
for researchers in other fields, especially in soft condensed matter physics.
We try to show that the pasta phases can be an interesting system
for many researchers in various fields, not only for nuclear astrophysicists.
The plan of this paper is as follows.
In the remaining part of the present section, we first give a brief
explanation of collapse-driven supernovae, neutron stars and
materials in these objects.
We then describe the basic physics and astrophysical consequences of the pasta phases.
In Section \ref{sect_soft}, we discuss the similarity between the pasta phases
and soft condensed matter.
We then explain a theoretical framework used in 
our studies in Section \ref{sect_md_nucleon},
and show the results in Section \ref{sect_result}.
In this article, we generally set the Boltzmann constant $k_{\rm B}=1$.

\subsection{Collapse-Driven Supernovae \label{SN explosion}}
Stars evolve by burning light elements into heavier ones
(see, e.g., Ref.\ \cite{whw}); 
here burning means nuclear fusion.
The ultimate fate of the star is basically
determined by its mass in the main sequence period, when the star is
supported by thermal pressure of burning hydrogen nuclei (protons).
Nuclear reactions of heavier nuclei require higher temperatures because
of their greater Coulomb barrier. 
In the core of massive stars with $M\gtrsim 8M_{\odot}$, the temperature
reaches a threshold to produce iron. Elements heavier than iron cannot be 
produced by nuclear fusion since the binding energy per nucleon
of iron is the greatest among all the elements.
Therefore iron nuclei, the major final product of 
a chain of the thermally driven nuclear reactions,
accumulate in the core of massive stars at the end of their evolution.

As the nuclear burning proceeds and the mass of the iron core increases,
two processes which tend to make the core unstable to collapse occur --- 
electron capture, and photodissociation of heavy nuclei.
In the iron core the density is so high
($ \sim 4\times 10^9 \mathrm{g}\,\mathrm{cm}^{-3}$)
that the core is supported mainly by electron degeneracy pressure;
degenerate electrons have extremely large momenta even at zero
temperature due to the Pauli exclusion principle.
When the electron Fermi energy exceeds
3.7 MeV, electron capture on iron nuclei occurs:
\begin{equation}
 ^{56}\mathrm{Fe}+\mathrm{e}^- \rightarrow\ ^{56}\mathrm{Mn}+\nu_{\rm e}\ .
 \label{electron_capture}
\end{equation}

Since this reaction decreases the electron density, the
electron degeneracy pressure decreases and can no longer 
support the iron core.
This instability leads to contraction of the core.
The other instability is triggered by
photodissociation of iron nuclei, which occurs when the temperature
of the core is $\gtrsim5\times 10^9$ K:
\begin{equation}
 \gamma +\ ^{56}_{26}\mathrm{Fe} \rightarrow 13\alpha + 4\mathrm{n}
 -124.4 \mathrm{MeV}\ .\label{photodissociation}
\end{equation}
This is an endothermic reaction; thus it decreases the gas pressure
and accelerates the collapse.

As the collapse of the core proceeds and the central density approaches
the nuclear saturation density (normal nuclear density),
$\rho_0=0.165 \mathrm{\ nucleons\ fm}^{-3}$
$\simeq 3\times 10^{14}\ \mathrm{g}\,\mathrm{cm}^{-3}$,
the equation of state suddenly becomes hard because of a strong short-range
repulsion between nucleons.
Due to this hardening, the pressure becomes sufficiently high to halt
the collapse, causing the inner region of the core to bounce.
The outer region of the core continues to fall towards the center at
supersonic velocities. Consequently, the bouncing inner core drives
a shock wave into the infalling outer core.
The initial energy of the shock is $\sim 10^{51}$ erg and is enough
to blow off the stellar envelope, which results in an explosion of the star.
However, the shock wave propagating through the outer core is
weakened by several processes (see, e.g., Ref.\ \cite{suzuki}) 
which decrease the pressure
and dissipate the energy behind the shock front. Consequently, the
shock wave stalls in the outer core.
Neutrinos emitted from the inner core will heat the matter behind
the shock front (a process referred to as neutrino heating). 
If the neutrino heating is efficient enough, 
the stalled shock can be revived,
reach the surface of the outer core, 
propagate beyond the core,
and finally blow off the outer layer of the star;
thereby producing a supernova explosion.
The contracted core remains as a nascent neutron star.

The difference in the gravitational energy between the iron core 
and a neutron star is given by
\begin{equation}
\Delta E \simeq -\left(
\frac{GM^2_{\mathrm{core}}}{R_{\mathrm{Fe}\,\,\mathrm{core}}}
-\frac{GM^2_{\mathrm{core}}}{R_{\mathrm{NS}}}
\right)
\simeq \frac{GM^2_{\mathrm{core}}}{R_{\mathrm{NS}}}
\sim O(10^{53})\ \mathrm{erg}\ ,\label{GravEnergyDiff}
\end{equation}
where $G$ is the gravitational constant,
$M_{\mathrm{core}} \sim 1 M_{\odot}$ is the mass of the core,
$R_{\mathrm{Fe}\,\,\mathrm{core}}\sim O(10^3)$ km is the
initial radius of the iron core 
and the neutron star radius $R_{\mathrm{NS}}$ is about 10 km.
Only one percent of this energy is injected into the gas blown off
and 99 \% is carried away by neutrinos
\footnote{The gravitational energy is emitted in
all six flavors of neutrinos and anti-neutrinos almost equally.
However, muon and tauon neutrinos (and anti-neutrinos) are not important
for supernova explosions because they interact with matter
extremely weakly. 
Thus, hereafter, the term ``neutrino'' denotes an electron neutrino.}.
In the supernova explosion, neutrinos interact with matter and will inject
kinetic energy into it. 
Therefore detailed investigations of
interactions between neutrinos and matter in supernova cores are
necessary to understand the explosion mechanism 
(see, e.g., Refs.\ \cite{bethe,burrows,fst review,janka,martinez}).

As the density in the core increases during the collapse, the mean
free path of neutrinos $l_{\nu}$
decreases mainly due to
the neutrino coherent scattering from nuclei via the weak neutral current.
The amplitude of the coherent scattering is proportional to
$A$, where $A$ is the mass number of the nucleus. Therefore its
cross section $\sigma$ is proportional to $A^2$ \cite{freedman},
while that of the incoherent scattering is simply
proportional to $A$.
When the neutrino wavelength is much longer than the radius of the
nucleus, the neutrino is coherently scattered
by nucleons in the nucleus. 
In the collapsing
iron core, a typical value of the wavelength $\lambda_{\nu}$ of a neutrino
with energy $E_{\nu}$ is
\begin{equation}
\lambda_{\nu}=\frac{hc}{E_{\nu}}\simeq 2\pi\times 20 \,\mathrm{fm}
\left(\frac{10\mathrm{MeV}}{E_{\nu}}\right) ,\label{mfp}
\end{equation}
and that of the nuclear radius $r_{\mathrm{N}}$ is
\begin{equation}
r_{\mathrm{N}}\simeq 1.2 A^{1/3} \,\mathrm{fm} \simeq 5 \,\mathrm{fm}
\left(\frac{A}{56}\right)^{1/3}.\label{nuclear radius}
\end{equation}

\begin{figure}[tbp]
\begin{center}
\rotatebox{0}{
\resizebox{8.5cm}{!}
{\includegraphics{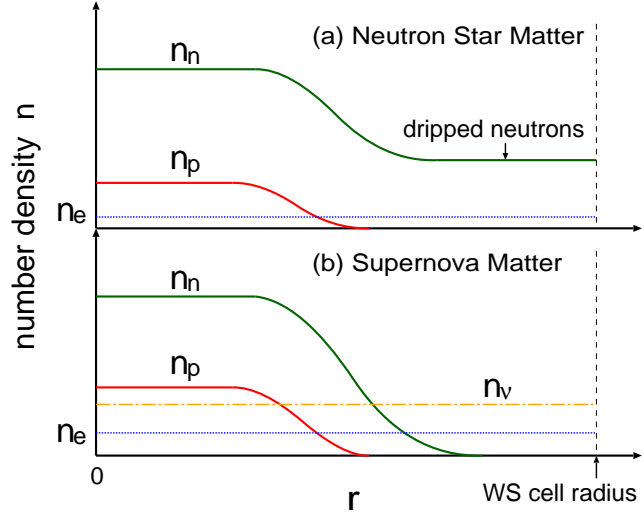}}}
\caption{\label{wscell}(Color)\quad
 Schematic diagram of a Wigner-Seitz cell of neutron star matter (NSM)
 and supernova matter (SNM) where $n_{\rm n}$, $n_{\rm p}$, $n_{\rm e}$ and $n_{\nu}$ are
 the number densities of neutrons, protons, electrons and neutrinos
 respectively.
  }
\end{center}
\end{figure}

The neutrino diffusion time scale is estimated by the random-walk
relation:
\begin{equation}
\tau_{\mathrm{diff}}\sim \frac{R^2}{c l_{\nu}}\ ,\label{diff time}
\end{equation}
where $R$ is the radius of the collapsing core and $c$ is 
the speed of light. 
The dynamical time scale of the core collapse is 
comparable to the free-fall time scale:
\begin{equation}
\tau_{\mathrm{dyn}}\sim \frac{1}{\sqrt{G\rho_{\mathrm{core}}}}\ ,
\label{dyn time}
\end{equation}
where $\rho_{\mathrm{core}}$ is the density of the core.
When $\tau_{\mathrm{diff}}$ is greater than $\tau_{\mathrm{dyn}}$ (the
corresponding density region is above
$10^{11-12}\mathrm{g}\,\mathrm{cm}^{-3}$), 
neutrinos cannot escape from the inner core \cite{sato}; 
this phenomenon is called neutrino trapping.
While neutrinos produced by electron capture remain in the inner core 
for the order of 10 msec,
the lepton fraction $Y_L$, the lepton
number per nucleon [$Y_L\equiv (n_{\rm e}+n_{\nu})/(n_{\rm n}+n_{\rm p})$ where
$n_{\rm e}$, $n_{\nu}$, $n_{\rm n}$ and $n_{\rm p}$ are the number densities of electrons,
neutrinos, neutrons and protons],
stays in a range of about 0.3 - 0.4.
Trapped neutrinos build up a degenerate sea, which suppresses
electron capture.
Because the time scale of weak interaction processes is much shorter than
$\tau_{\mathrm{dyn}}\sim\tau_{\mathrm{diff}}$, matter in the inner core
is in $\beta$ equilibrium with trapped neutrinos.
At densities $\rho\lesssim\rho_0$, matter consists of
nuclei, electrons (to maintain charge neutrality) and neutrinos,
and we shall refer to this simply as supernova matter (SNM) \cite{bonche,lamb}.
We show particle number density profiles in supernova matter 
in Fig.\ \ref{wscell}-(b).
Since degenerate electrons are relativistic, screening effects are
negligible and the electrons are uniformly distributed
\cite{maruyama_screen,review,screening}.
There also exist the uniform neutrino gas
and neutron-rich nuclei: clusters of protons and neutrons.
In a certain density region below the normal nuclear density, 
supernova matter could consist
of nuclei with exotic shapes such as rod-like and slab-like nuclei
rather than spherical ones, which are referred to as nuclear ``pasta''
(a further explanation will be given in Section \ref{sect_what}).
The pasta nuclei affect neutrino opacity of the supernova matter
(see Section \ref{astro conseq}).

\subsection{Neutron Stars \label{NS structure}}

Let us see further evolution of the bounced core.
After bounce, the core settles into hydrostatic equilibrium on its dynamical
time scale and a protoneutron star is formed unless the core is not so heavy
that it collapses to a black hole.
The initial radius and temperature of the protoneutron star
are $\sim 100$ km and $\sim 10$ MeV, respectively.
Just after the bounce of the core, the proton fraction 
$x\equiv n_{\rm p}/(n_{\rm n}+n_{\rm p})$,
and the lepton fraction of matter in the core 
are relatively high (both are around 0.3).
As neutrinos escape from the protoneutron star, it cools down
and electron capture, which is blocked by the degenerate neutrinos, proceeds (see, e.g., Ref.\ \cite{yakovlev} 
for a recent review of cooling of neutron stars).
Consequently matter gets neutron rich; the proton
fraction $x\lesssim 0.1$ at around the normal nuclear density
$\rho_0$.
At the same time, the radius of the object shrinks to $\sim 10$ km; a neutron
star is formed.

Neutron stars are dense and compact objects supported by neutron degeneracy
pressure and nuclear forces. The mass and the radius of a typical neutron star are
$\simeq 1.4 M_{\odot}$ and $\simeq 10\ \mathrm{km}$, respectively.
The properties of dense matter and its
equation of state are relatively well understood in the density region below
$\sim \rho_0$.
Thus the theoretical picture of neutron star structure is reasonably
established in the lower density region
(see, e.g., Ref.\ \cite{lattimer,review} for reviews).

\begin{figure}[t]
\begin{center}
\rotatebox{0}{
\resizebox{11cm}{!}
{\includegraphics{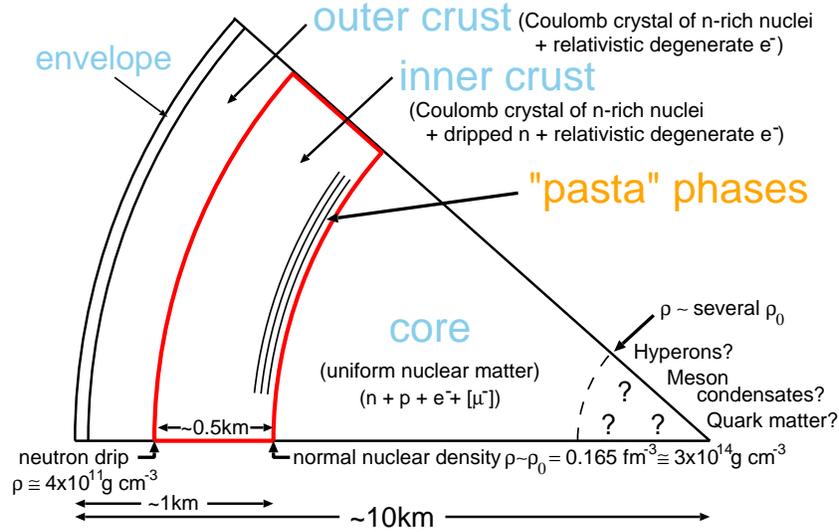}}}
\caption{\label{cross section of NS}(Color)\quad
 Schematic picture of the cross section of a neutron star.
  }
\end{center}
\end{figure}

In Fig.\ \ref{cross section of NS} 
we show a schematic cross section of a typical cold neutron star
below $10^9$ K.
The outermost part, the envelope, is a layer about
several meters thick which consists of liquid $^{56}\mathrm{Fe}$
(if there is accretion from its companion star, the envelope is covered
by a layer of hydrogen and helium atoms).
With increasing density 
(i.e., proceeding to the deeper region),
weak interactions render nuclei neutron-rich via electron captures
triggered by a large electron chemical potential. Matter becomes 
solid at a density $\rho \gtrsim 10^6 \mathrm{g}\,\mathrm{cm^{-3}}$,
which consists of nuclei forming a bcc Coulomb lattice neutralized
by a roughly uniform electrons. Then, at a density of about 
$\sim 4 \times 10^{11} \mathrm{g}\,\mathrm{cm^{-3}}$, nuclei
become so neutron rich that the last occupied neutron levels are
no longer bound; neutrons begin to drip out of these nuclei.
These ``dripped'' neutrons form a superfluid amid the neutron-rich
nuclei.
The crystalline region of the star is referred to as crust,
which is divided into inner and outer crust;
in the outer crust there are no neutrons outside nuclei, 
while in the inner crust there are.
The inner crust extends from the neutron drip point
to the boundary with the core at a density 
$\rho \lesssim \rho_0 \simeq 3 \times 10^{14}\mathrm{g}\,\mathrm{cm^{-3}}$,
where nuclei disappear and the system becomes uniform
nuclear matter.
Matter in these regions is called neutron star matter (NSM)
\cite{bbp,bbs,negele}.
A schematic picture of neutron star matter at subnuclear densities is shown in Fig.\ \ref{wscell}-(a).
There exist dripped neutrons outside the neutron-rich nuclei.
Degenerate electrons are uniformly distributed for the same reason
as in the case of supernova matter. 
Another difference between supernova matter and 
neutron star matter is the lack of trapped degenerate neutrinos. 
This makes neutron star matter more neutron-rich
than supernova matter in $\beta$ equilibrium.
In the deepest 
region of the inner crust
corresponding to subnuclear densities (i.e., $\rho\lesssim\rho_0$), 
pasta nuclei could appear (see Section \ref{sect_what}
for a further explanation).
Although crusts of neutron stars are relatively thin ($\sim 1$ km), 
they influence many observed phenomena \cite{lattimer,review}.
Effects of the pasta phases on these phenomena will be discussed
in Section \ref{astro conseq}.

The physics of the density region above several times normal nuclear density is quite
uncertain, and a variety of hadronic phases have been proposed, such as
hyperonic matter, pion or kaon condensates, quark-hadron mixed phase
and uniform quark matter, etc.
(see, e.g., Ref.\ \cite{heiselberg review} and references therein for details).
These are beyond the scope of the present article.

\subsection{What is Nuclear ``Pasta''? \label{sect_what}}

\begin{figure}[ht]
\begin{center}
\rotatebox{0}{
\resizebox{10cm}{!}
{\includegraphics{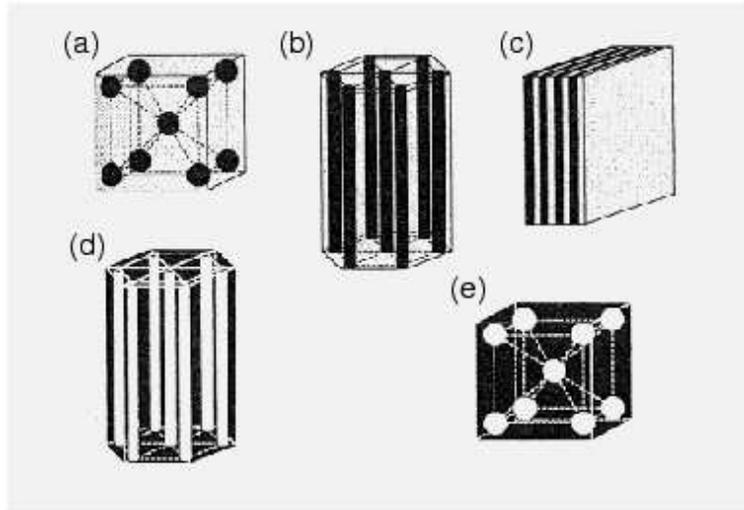}}}
\caption{\label{pasta}
  Nuclear ``pasta.''
  The darker regions show the liquid phase,
  in which protons and neutrons coexist (i.e., the nuclear matter region);
  the lighter ones the gas phase, which is almost free of protons.
  Sequence (a)-(e) shows that of nuclear shape changes
  with increasing density.
  This figure is taken from Ref.\ \cite{oyamatsu}.}
\end{center}
\end{figure}

In ordinary matter, atomic nuclei are roughly spherical.
This may be understood in the liquid drop picture of the nucleus as being a 
result of the forces due to the surface tension of nuclear matter,
which favors a spherical nucleus, being greater than those due to the 
electrical repulsion between protons, which tends to make the nucleus deform.
When the density of matter approaches that of atomic nuclei, i.e.,
the normal nuclear density $\rho_0$, nuclei are closely packed
and the effect of the electrostatic energy becomes comparable
to that of the surface energy.
Consequently, at subnuclear densities around $\rho\lesssim\rho_0/2$,
the energetically favorable configuration is expected to have
remarkable structures as shown in Fig.\ \ref{pasta};
the nuclear matter region (i.e., the liquid phase of mixture of protons and neutrons)
is divided into periodically arranged parts of roughly
spherical (a), rod-like (b) or slab-like (c) shape, embedded in the gas phase and
in a roughly uniform electron gas.
Besides, there can be phases in which nuclei are turned inside out,
with cylindrical (d) or spherical (e) bubbles of the gas phase
in the liquid phase.
As mentioned in the previous section,
these transformations are expected to occur 
in the deepest region of neutron
star inner crusts and in the inner cores of collapsing stars
just before the star rebounds.
Since slabs and rods look like ``lasagna'' and ``spaghetti'',
the phases with non-spherical nuclei
are often referred to as ``pasta'' phases and
such non-spherical nuclei as nuclear ``pasta.''
Likewise, spherical nuclei and spherical bubbles are called
``meatballs'' and ``Swiss cheese'', respectively.

More than twenty years ago,
Ravenhall {\it et al.} \cite{rpw} and Hashimoto {\it et al.} \cite{hashimoto}
independently pointed out that nuclei with such exotic shapes
\footnote{The ``Swiss cheese'' phase was considered earlier in Ref.\ \cite{bbp}
than the other pasta phases.} can be
the most energetically stable
due to the subtle competition between the nuclear surface and Coulomb energies
as mentioned above.
Let us here show this statement by a simple calculation
using an incompressible liquid-drop model
(see also Ref.\ \cite{oyamatsu phd}).

We consider five phases depicted in Fig.\ \ref{pasta},
which consist of spherical nuclei, cylindrical nuclei,
planar nuclei, cylindrical bubbles and spherical bubbles, respectively.
Each phase is taken to be composed of a single species of 
nucleus or bubble at a given nucleon density $\rho$ averaged over the space.
The total electrostatic energy including the lattice energy,
which is the contribution to the Coulomb energy 
coming from neighbouring nuclei,
is estimated by the Wigner-Seitz approximation.

In the present discussion,
we assume that the nuclear matter region is incompressible
and the nucleon density there is fixed at the normal nuclear density
$\rho_{0}=0.165$ fm$^{-3}$.
We further assume that the proton fraction $x_{\rm N}$
in the nuclear matter region is fixed at $x_{\rm N}=0.3$
and that the dripped neutrons can be ignored for simplicity.
These assumptions correspond to the situation of supernova matter
[see Fig.\ \ref{wscell}-(b)].
Using a liquid-drop model, the total energy density of matter $E_{\rm tot}$
averaged over a single cell can be written as
\begin{equation}
  E_{\rm tot}=\left\{
    \begin{array}{ll}
      uw_{\rm bulk}+w_{\rm surf}+w_{\rm C+L}+E_{\rm e}& \mbox{(nuclei)}\ , \\
      (1-u)w_{\rm bulk}+w_{\rm surf}+w_{\rm C+L}+E_{\rm e}&\mbox{(bubbles)}\ ,
    \end{array}\right.\label{total energy density}
\end{equation}
where $u$ is the volume fraction of the nuclei or bubbles,
$w_{\rm bulk}$ is the bulk energy density of the nuclear matter region,
$w_{\rm surf},\ w_{\rm C+L}$, and $E_{\rm e}$ are the nuclear surface,
total electrostatic, and electron energy densities, respectively.
Under the assumption of neglecting the dripped neutrons,
the volume fraction $u$ of nuclei [bubbles] is given as
$u=\rho/\rho_{0}$ $[u=1-(\rho/\rho_{0})]$.
The electron number density $n_{\rm e}$, which is assumed to be uniform here,
is $n_{\rm e}=\rho_{0}x_{\rm N}u$ $[n_{\rm e}=\rho_{0}x_{\rm N}(1-u)]$.
As we consider incompressible nuclear matter
with fixed proton fraction $x_{\rm N}$,
in the expression of $E_{\rm tot}$ [Eq.\ (\ref{total energy density})]
only $w_{\rm surf}$ and $w_{\rm C+L}$ depend on the nuclear shape
at a given $\rho$.
Thus the energetically most stable nuclear shape can be determined by
comparing $w_{\rm surf}+w_{\rm C+L}$ between each nuclear shape.

The surface energy per unit volume $w_{\rm surf}$ is expressed as
\begin{equation}
  w_{\rm surf}=\frac{\sigma ud}{r_{\rm N}}\ ,
\end{equation}
where $\sigma$ is the surface tension,
$r_{\rm N}$ is the radius (or the half width) of nuclei or bubbles
and $d$ is the dimensionality of the nuclear shape defined as
$d=1$ for slabs,
$d=2$ for cylinders or cylindrical holes and,
$d=3$ for spheres or spherical holes.
We assume $\sigma$ to be constant and
set $\sigma=0.73$ MeV fm$^{-2}$ \cite{rbp}.

In the Wigner-Seitz approximation, one replaces the actual unit cell
by a spherical one for $d=3$, a cylindrical one for $d=2$,
and a planar one for $d=1$, which are electrically neutral and have
an equal volume to the actual cell. Using this approximation,
the total electrostatic energy per unit volume $w_{\rm C+L}$ yields
\begin{equation}
w_{\rm C+L}=2\pi(\mathrm{e}x_{\rm N}\rho_{0} r_{\rm N})^2 u f_d(u)\ ,
\label{coulomb}
\end{equation}
with
\begin{equation}
f_d(u)= \frac{1}{d+2}
\left[\frac{2}{d-2} \left( 1-\frac{d u^{1-2/d}}{2} \right) +u \right]\ .
\end{equation}

Minimizing $w_{\rm surf}+w_{\rm C+L}$ with respect to $r_{\rm N}$
leads to a simple condition for all the five nuclear shapes considered here:
\begin{equation}
  w_{\rm surf}=2w_{\rm C+L}\ .\label{size eq}
\end{equation}
Thus $w_{\rm surf}+w_{\rm C+L}$ is given as
\begin{equation}
  w_{\rm surf}+w_{\rm C+L}=\rho_{0}\zeta_{s}g_{d}(u)\ ,\label{surf+coulomb}
\end{equation}
with
$\zeta_{s}\equiv 3
\left(\frac{9}{5} \pi x_{\rm N}^{2}\mathrm{e}^{2}\sigma^{2} \frac{1}{\rho_{0}}
\right)^{1/3}$
and $g_{d}(u) \equiv u \left( \frac{5}{18} d^{2} f_{d}(u) \right)^{1/3}$.


\begin{figure}[htbp]
\centering
\includegraphics[width=0.45\textwidth]{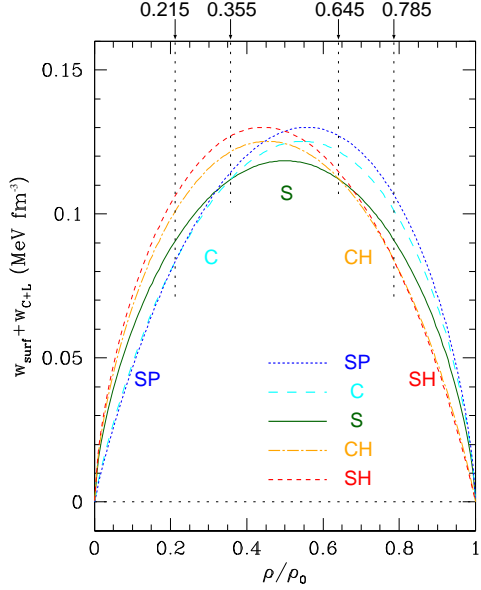}
\caption{(Color)\quad
  $w_{\rm surf}+w_{\rm C+L}$ given by Eq.\ (\ref{surf+coulomb})
  as a function of the nucleon number density $\rho$.
  The symbols SP, C, S, CH, and SH stand for nuclear shapes,
  i.e., sphere, cylinder, slab, cylindrical hole, and spherical hole,
  respectively.
  The transition points locate symmetrically with respect to the point of
  $\rho=\rho_{0}/2$ in the case of neglecting
  the bulk contribution.}
\label{fig surf+coulomb}
\end{figure}

We have plotted in Fig.\ \ref{fig surf+coulomb}
the summation of the nuclear surface and total electrostatic energy densities,
$w_{\rm surf}+w_{\rm C+L}$, given by Eq.\ (\ref{surf+coulomb})
for each five nuclear shape.
At densities below $\rho_{0}$, the system must have an inhomogeneous structure
of some kind because matter is assumed to be incompressible here.
The point at $\rho=\rho_{0}$,
where the lines for all shapes and
the line of $w_{\rm surf}+w_{\rm C+L}=0$ cross each other,
corresponds to uniform nuclear matter.
It can be seen from Fig.\ \ref{fig surf+coulomb} that
the phase with spherical nuclei gives the lowest value of
$w_{\rm surf}+w_{\rm C+L}$ at lower densities below 0.215 $\rho_0$,
and, with increasing density,
the most stable shape of the nuclear matter region changes as
\begin{equation}
  \mbox{sphere} \rightarrow \mbox{cylinder} \rightarrow \mbox{slab}
  \rightarrow \mbox{cylindrical hole} \rightarrow \mbox{spherical hole}
  \rightarrow \mbox{uniform}\ .\label{sequence}
\end{equation}

Even though the pasta phases were favored energetically,
whether or not they are actually present in the star
depends on occurrence of the dynamical processes leading to their formation.
Our recent dynamical simulations \cite{qmd_transition,qmd1,qmd2,qmd_hot} 
have almost solved this problem (see Section \ref{sect_result}).

\subsection{Astrophysical Consequences \label{astro conseq}}

The presence of nuclear pasta would affect some astrophysical
phenomena. 
It is noted that the pasta phases can occupy half of the total mass of the
neutron star crusts \cite{lorenz} and 10 -- 20\% of the mass of supernova cores
in the later stage of the collapse \cite{sonoda}.
In this section we explain effects of the pasta phases
on the neutrino trapping and
the core dynamics of supernovae, pulsar glitches and cooling of
neutron stars.

\subsubsection{Supernova Explosions\label{SNexp}}

\begin{figure}[t]
\begin{center}
\rotatebox{0}{
\resizebox{8cm}{!}
{\includegraphics{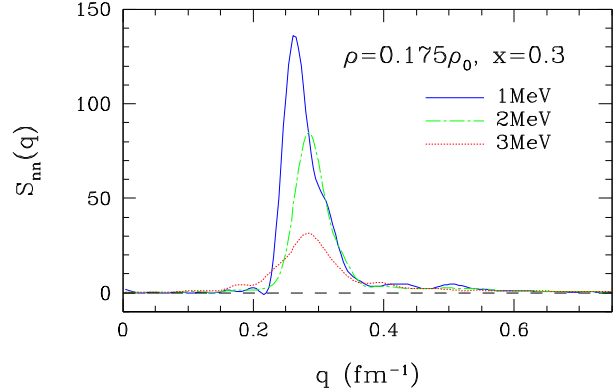}}}
\caption{\label{sq}(Color)\quad
 Static structure factor of neutrons $S_{\rm nn}(q)$ calculated for nucleon
 distributions at $x=0.3$, $\rho=0.175\ \rho_0$
 (rod phase at low temperatures),
 and at temperatures of 1MeV (solid line), 2MeV
 (dash-dotted line) and 3MeV (dotted line).
  }
\end{center}
\end{figure}

Formation of the pasta phases in collapsing cores would
influence neutrino transport in the cores.
As mentioned in Section \ref{SN explosion}, this is
considered to be crucial for reproduction of supernova explosions
by simulation studies.
Due to neutrino-nucleus coherent scattering, 
the pasta phases have a large effect on the neutrino opacity
of supernova matter \cite{gentaro2,qmd1,qmd2}.
If one takes account of the pasta phases, they replace uniform matter
close to a boundary between uniform and non-uniform phases;
the neutrino opacity of the pasta phases is significantly larger
than that of uniform nuclear matter. 
This enhancement is elaborated with quantum molecular dynamic (QMD) 
simulations in Refs.\ \cite{horowitz2,horowitz3,sonoda} 
(see also Section \ref{sect_md_nucleon} for the framework of the QMD).
The cross section for coherent neutrino scattering is approximated to be
proportional to the static structure factor $S_{\rm nn}(q)$ of neutrons,
where $q$ is the wave number of the momentum transfer.
The static structure factor is obtained by the Fourier transform of
a radial distribution function which represents two-point correlations.
This quantity is remarkably
increased near the wave number corresponding to the
reciprocal lattice vector; in the case of uniform matter this peak 
cannot be seen. 
We show in Fig.\ \ref{sq} the static structure factor $S_{\rm nn}(q)$
calculated for nucleon distributions of $T=1$, 2 and 3 MeV
at $x=0.3$ and $\rho=0.175$ $\rho_0$, where the phase with rod-like nuclei appears
at low temperatures. 
The nucleon distributions are obtained by our QMD simulations,
which will be explained in Section \ref{sect_result}.
A peak around $q=0.25$ -- 0.3 fm$^{-1}$ at each temperature indicates
enhancement of the neutrino opacity; at 1 MeV the cross section
around the peak exceeds one hundred times that for
uniform matter.
With increasing temperature, the peak height
drops because nuclei start to melt 
(see phase diagrams in Section \ref{sect_result}).

The existence of pasta nuclei also affects the hydrodynamics of
supernova cores.
Phase transition from a bcc lattice to uniform nuclear matter
is considerably smoothed by the presence of the pasta phases,
which modifies the equation of state used in core collapse simulations
\cite{rpw}.
In addition to the equation of state, the pasta phases also have an effect on
the other hydrodynamic properties 
of supernova matter due to their different elastic properties from
those of a crystalline lattice of spherical nuclei or uniform nuclear matter.
Although the above effects are crucial for supernova explosions,
it is not obvious how they affect the explosions in detail; this should be
investigated by simulation studies of supernovae in the future.

\subsubsection{Pulsar Glitches\label{glitches}}
Pulsars are neutron stars emitting periodic pulses, with a period 
equal to that of the rotation of the star.
In more than 20 pulsars, sudden decreases of pulse periods, i.e., 
spinups of neutron star crusts, are observed. These phenomena are
called ``glitches.'' There is not yet a consensus about the mechanism
of glitches but a widely accepted theory is the vortex pinning model
\cite{anderson}.

In neutron star crusts, dripped neutrons form a superfluid and
neutron-rich nuclei, which may be regarded as a normal fluid, act as a pinning center
for vortex lines in the neutron superfluid.
In the vortex pinning model glitches represent vorticity jumps
following catastrophic unpinning of the vortex lines.
Since migrations of vortex lines result in transport of angular momentum,
this leads to spinups of the crusts.
It is obvious that the existence of the nuclear pasta would have
a large influence on pinning.
However, the force needed to pin the vortices and the pinning rate
have yet to be clarified completely even for a bcc lattice of
spherical nuclei, rather than for non-spherical nuclei \cite{jones}.

There is another representative model of pulsar
glitches: the starquake model \cite{starquake}.
According to this model, a sudden starquake of the neutron star
crust decreases the moment of inertia and hence increases the angular velocity.
Nuclei in the crust form a solid, Coulomb lattice. The rigidity
of the solid crust restores the force against deformation.
When the stress in the crust reaches a critical value, the crust cracks.
The existence of the pasta phases instead of the normal bcc lattice
reduces the maximum elastic energy that can be stored in the crust
because, as will be discussed in Section \ref{sect_soft},
the pasta phases have directions in which restoring force
does not act like liquid crystals \cite{pp}.
Thus the pasta phases would reduce the strength of each
glitch and shorten the time separation between glitches in this
scenario.

\subsubsection{Cooling of Neutron Stars\label{cooling of NS}}

In the early epoch of the neutron star cooling, the main
process of energy loss by neutrinos is the so-called
URCA process:
\begin{equation}
\mathrm{n}\rightarrow \mathrm{p}+\mathrm{e}^-+\bar{\nu}_{\rm e}\,\,\,\mathrm{and}\,\,\,
\mathrm{p}+\mathrm{e}^-\rightarrow \mathrm{n}+\nu_{\rm e}\ . \label{dURCA}
\end{equation}
These reactions are dominant in the case that matter in the neutron
star is hot and non-degenerate. However, when the matter becomes cold
and degenerate as in a neutron star, the URCA process is strongly
suppressed by the energy and momentum conservation laws.

The presence of non-spherical nuclei would accelerate the cooling
of neutron stars by opening the direct URCA reactions which are
unlikely to occur for spherical nuclei \cite{lorenz}.
This stems from the fact that, in non-spherical nuclei, protons
have a continuous spectrum at the Fermi surface in the elongated
directions.

The following reactions are generally accepted as the mechanism of
generating neutrinos in cold and degenerate neutron stars:
\begin{equation}
\mathrm{n}+\mathrm{n}\rightarrow \mathrm{n}+\mathrm{p}+\mathrm{e}^-+\bar{\nu}_{\rm e}\,\,\,\mathrm{and}\,\,\,
\mathrm{n}+\mathrm{p}+\mathrm{e}^-\rightarrow \mathrm{n}+\mathrm{n}+\nu_{\rm e}\ , \label{mURCA}
\end{equation}
which is referred to as the modified URCA process, and 
neutrino pair bremsstrahlung by electrons in the crust:
\begin{equation}
\mathrm{e}^-+(Z,A)\rightarrow \mathrm{e}^- +(Z,A)+\nu+\bar{\nu}\ .\label{brems}
\end{equation}
In the modified URCA process, a bystander neutron absorbs the momentum
and relaxes kinematical constraints.

Pasta nuclei suppress the bremsstrahlung rate
\cite{kaminker,thorsson}. If we assume that neutrino bremsstrahlung is
dominated by the lowest reciprocal lattice vectors for which form
factors do not vanish, the number of these vectors for each dimensional
lattice leads to the
neutrino emission rate of spherical, cylindrical and planar nuclei
in the ratio of 6:3:1.
If the nucleons in the core
undergo a transition to a superfluid and/or superconducting state,
the modified URCA process is suppressed by a factor of
$\sim e^{-\Delta/k_{\rm B} T}$, where $\Delta$ is the superfluid energy gap.
In this case bremsstrahlung is dominant and
the neutrino emission is suppressed efficiently as a
consequence of the presence of pasta phases.

\section{Pasta Phases as Soft Condensed Matter\label{sect_soft}}

Nuclear systems can be regarded as complex fluids of nucleons.
Complex fluids are the systems in which scales of the fluid 
and their constituents cannot be separated clearly;
both scales are strongly connected with each other.
Such characteristic can be seen in various soft condensed matter systems.
For example, polymer systems are one of the typical complex fluids
(description of polymer systems in the present section
is mainly based on an instructive review \cite{kawakatsu}).
Actually, similar spatial structures to nuclear pasta 
as shown in Fig.\ \ref{pasta} can be seen in some polymer systems
such as domain structures in solutions and melts of block copolymers.
Thus it would be interesting to compare the hierarchical structure
of nuclear systems with that of polymer systems.

\begin{figure}
\begin{center}
\rotatebox{270}{
\resizebox{6.5cm}{!}
{\includegraphics{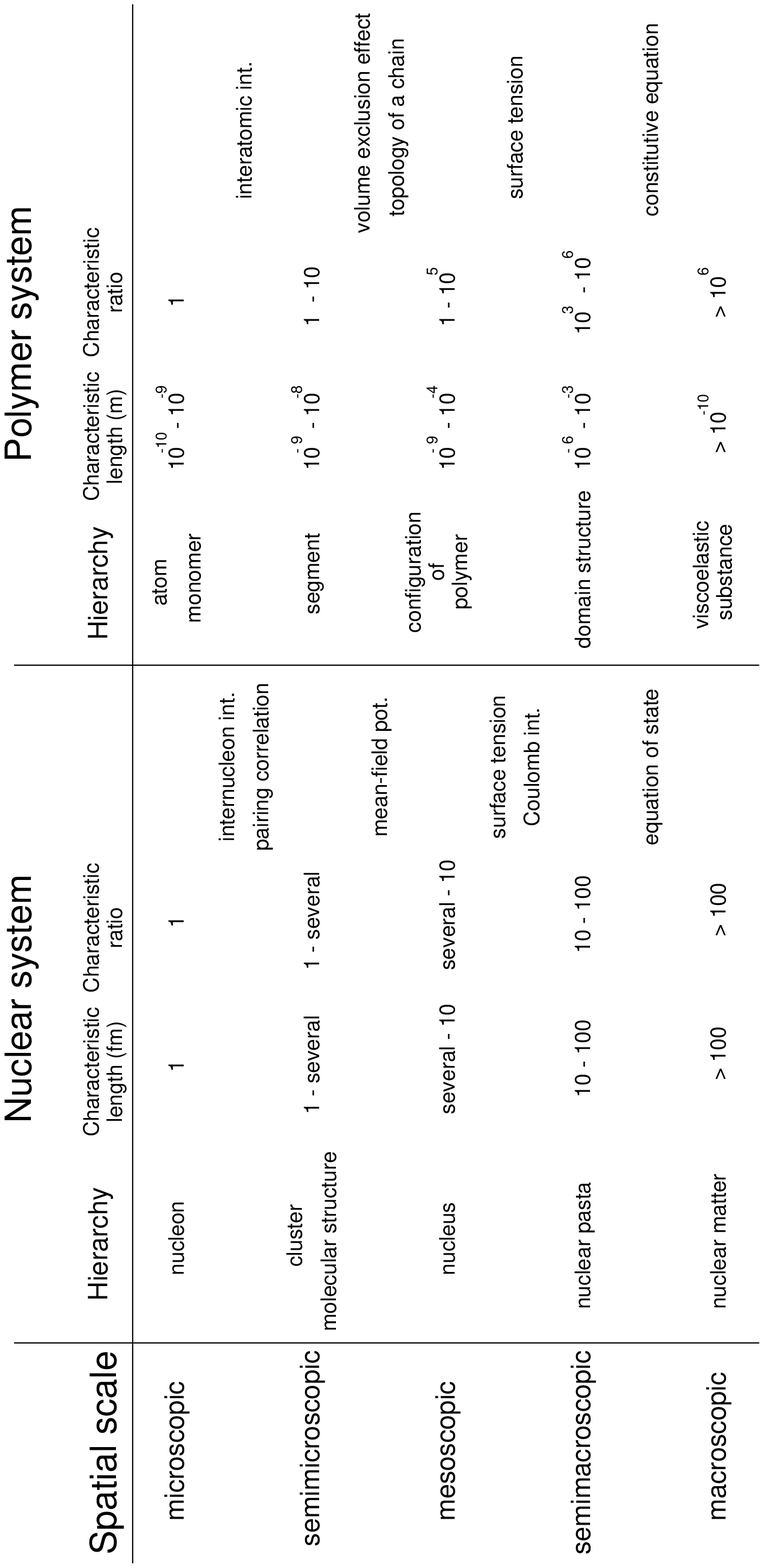}}}
\caption{\label{fig hierarchy}
  Hierarchical structures of nuclear system and of polymer system.
  We refer to Ref.\ \cite{kawakatsu} for polymer system.
  }
\end{center}
\end{figure}

Polymers are made up of monomers connected by covalent bondings
and their degrees of polymerization are quite large ($> 100$).
Consequently, in polymer systems, the scale of constituents is
as large as that of macroscopic phenomena such as 
phase separation of the domain structures.
Thus polymer systems behave as complex fluids.
In nuclear systems, the nature of a complex fluid
is caused by strong and short range nuclear force.

In Fig.\ \ref{fig hierarchy}, we show the hierarchical structures
of nuclear systems and of polymer systems.
Nuclear pasta phases and domain structures of block copolymers
can be classified in the semimacroscopic scale. Thus theoretical
methods based on the mesoscopic or semimacroscopic scales are practical
for studying these systems.
In the study of polymer systems, the density functional theory and
the Ginzburg-Landau theory correspond to such approaches.
In nuclear physics, we have the Thomas-Fermi theory and the liquid-drop model,
which are based on the scale of the nucleon density profile and of the nucleus,
respectively.
Actually, thermodynamic properties of the pasta phases and
phase diagrams at zero-temperature have been investigated so far
mainly by these approaches.

However, the nature of complex fluids of nuclear systems
sets limits on the validity of the above methods.
They cannot be used for investigating phenomena 
in which microscopic degrees of freedom are important:
formation process of the pasta phases,
thermal fluctuations and melting of the pasta structures are typical examples.
However, due to the short range nuclear interaction,
characteristic ratio of the macroscopic or semimacroscopic scale 
in nuclear systems is much smaller than that of polymer systems 
(see Fig.\ \ref{fig hierarchy}).
Such a small characteristic ratio enables us to study the pasta phases
by microscopic approaches based on the nucleonic degrees of freedom.
Our simulations are performed using one of these methods,
which will be explained in the next section.

Similarity between the pasta phases and some phases of liquid crystals
should be also mentioned here.
For the pasta phases with rod-like nuclei and slab-like nuclei, 
there are directions in which the system is translationally invariant.
Consequently, any restoring forces do not act along such directions.
As noted by Pethick and his coworkers \cite{pp,review}, this situation is 
analogous with a liquid crystal rather than with a rigid solid. 
Elastic properties of the phases with rod-like nuclei and with slab-like ones
are thus characterized by elastic constants 
used for the corresponding liquid crystal phases, i.e., 
columnar phases and smectics A, respectively.

For the layered phases such as smectic A liquid crystals
and the pasta phase with slab-like nuclei,
the energy density $E_{\rm el}$ due to displacement $\epsilon$ of layers 
in their normal direction (taken to be parallel to the $z$-axis) 
can be written as \cite{degennes}
\begin{equation}
  E_{\rm el}\simeq 
  \frac{B}{2}\left(\frac{\partial \epsilon}{\partial z}\right)^2
  + \frac{K_1}{2}(\nabla_{\perp}^2 \epsilon)^2 +O(\epsilon^3)\ ,
\label{elast_slab}
\end{equation}
where $\nabla_{\perp}$ shows the derivative on the $x$-$y$ plane.
The elastic constants $B$ and $K_1$ are associated with
change of the layer spacing and with bending of the slabs.

For the two-dimensional triangular lattice
in the columnar phase of liquid crystals
or in the pasta phase with rod-like nuclei, 
the energy density due to a two-dimensional displacement vector 
$\mbox{\boldmath $\epsilon$}=(\epsilon_x, \epsilon_y)$ of rods
running along the $z$-axis is
given by \cite{degennes}
\begin{equation}
  E_{\rm el}\simeq \frac{B}{2}(\nabla_{\perp}\cdot
  \mbox{\boldmath $\epsilon$})^2
  +\frac{C}{2}\left[\left(\frac{\partial \epsilon_x}{\partial x}
  -\frac{\partial \epsilon_y}{\partial y}\right)^2
  +\left(\frac{\partial \epsilon_x}{\partial y} 
  + \frac{\partial \epsilon_y}{\partial x}
  \right)^2 \right] + \frac{K_3}{2}\left(
  \frac{\partial^2 \mbox{\boldmath $\epsilon$}}{\partial z^2}\right)^2
  +O(\epsilon^3)\ .\label{elast_rod}
\end{equation}
The elastic constant $B$ is associated with uniform translational compression
or dilation, $C$ with transverse shear and $K_3$ with bending of the rods.

Now let us discuss thermodynamic stability of the phase with slab-like nuclei.
According to the Landau-Peierls argument, strict long range positional order of
layered phases will be destroyed at finite temperatures \cite{landau}.
Physically, the elastic constant $B$ for compression of the spacing
between slab-like nuclei should be
comparable to the Coulomb energy density:
\begin{equation}
  B\sim w_{\rm C+L}\ ,
\end{equation}
and the elastic constant for bending of slab-like nuclei $K_1$ should 
have dependence of the surface energy density times squared of 
the internuclear spacing $l$:
\begin{equation}
  K_1\sim w_{\rm surf}\ l^2 \sim w_{\rm C+L}\ l^2\ ,
\end{equation}
here we use the fact that $w_{\rm surf}$ and $w_{\rm C+L}$ are comparable
in the equilibrium state [Eq.\ (\ref{size eq})].
The above expressions are consistent with results of an explicit calculation
in Ref.\ \cite{pp}.

For the one-dimensional layered lattice, 
the mean-square displacement can then be evaluated within the harmonic
approximation as \cite{chandra}
\begin{equation}
  \langle \epsilon^2 \rangle 
  \sim \frac{T}{\sqrt{B K_1}} \ln{\left(\frac{L}{l}\right)}
  \sim \frac{T}{w_{\rm C+L}\ l} \ln{\left(\frac{L}{l}\right)}\ ,
\label{msd slab}
\end{equation}
where $L$ is the one-dimensional length scale of the extension of the slab.
The Landau-Peierls instability for an infinite one-dimensional layered lattice 
can be seen by the logarithmic divergence of $\langle \epsilon^2 \rangle$
for large $L$.

Typical value of the interlayer spacing is $l\simeq 20$ fm.
As can be seen from Fig.\ \ref{fig surf+coulomb},
$w_{\rm C+L}$ in the supernova matter is $\sim 10^{-1}$ MeV fm$^{-3}$.
Employing the Lindemann criterion $\langle \epsilon^2 \rangle/l^2 \sim 0.1$
for melting of the lattice, limits of the extension $L$ are
given by Eq.\ (\ref{msd slab}) as $\ln{(L/l)}\lesssim 100$ 
at $T\simeq1$ MeV
and $\ln{(L/l)}\lesssim 10$ at $T\simeq10$ MeV.
In the initial phase of the stellar collapse before the bounce of the core,
the temperature of the supernova matter is the order of 1 MeV and
the limit of $L$ is even much larger than the radius of the core.
Thus the Landau-Peierls instability does not matter in this stage.

In the neutron star inner crusts, temperature is $\sim 0.1$ MeV.
The proton fraction in nuclei $x_{\rm N}$ is about one-third smaller
than that for supernova matter; as can be seen from Eq.\ (\ref{coulomb}),
$w_{\rm C+L}$ in neutron star matter is thus 
$w_{\rm C+L}\sim 10^{-2}$ MeV fm$^{-3}$ 
[other relevant quantities are comparable in the both cases].
Following the same way as in the above case, the limit of $L$ is given by
$\ln{(L/l)}\lesssim 100$, which is even much larger than 
the radius of neutron stars.
The phase with slab-like nuclei in neutron star crusts is thus stable 
in reality with respective to the Landau-Peierls instability.

Let us then estimate the temperature 
at which the lattice of the rod-like nuclei melts.
For the two-dimensional triangular lattice, 
the mean-square displacement is \cite{chandra}
\begin{equation}
  \langle \epsilon^2 \rangle \sim \frac{T}{(B+2C)\sqrt{\lambda l}}\ ,
\label{msd rod}
\end{equation}
where $\lambda\equiv\sqrt{2K_3/(B+2C)}$.
An explicit calculation in Ref.\ \cite{pp} shows that 
values of the elastic constants are
$B\sim w_{\rm C+L}$, $C\simeq w_{\rm C+L}$ and 
$K_3 \simeq 0.05\ w_{\rm C+L}\ l^2$.
Using the Lindemann criterion and the above typical values for 
$l$ and $w_{\rm C+L}$, one can estimate the melting temperature
$T_{\rm m} \gtrsim 10$ MeV for supernova matter and 
$T_{\rm m} \gtrsim 1$ MeV
for neutron star matter.
In the above estimate, however, decrease of the surface tension due to the
evaporation of nucleons and thermal broadening of the nuclear density profile
is not taken into account. Thus the melting temperature for supernova matter
would be smaller than the above value in the real situation.
More elaborated calculations on the Landau-Peierls instability
and of the melting temperature can be seen 
in Refs.\ \cite{olsson,gentaro1,gentaro2}.

In closing the present section, let us briefly mention 
several phases which are not shown in Fig.\ \ref{pasta}.
In block copolymer melts, various phases with complicated structures, e.g.,
gyroid phase, perforated lamellar phase and double diamond phase, etc.,
have been observed. As will be seen in Section \ref{sect_result},
our simulations suggest that phases with multiply connected structures
also appear in the nuclear systems (see also Ref.\ \cite{review}).
This point should be examined in the future study
\footnote{In the nuclear pasta phases, the long range Coulomb interaction
plays an essential role. 
Thus a Helfrich-type
simple local theory for the nuclear surface cannot be applied
to the present problem.
}.

\section{Molecular Dynamics for Nucleons\label{sect_md_nucleon}}

Since the seminal works by Ravenhall {\it et al.} \cite{rpw} and
Hashimoto {\it et al.} \cite{hashimoto}, properties of the pasta phases 
in equilibrium states have been investigated using various nuclear models.
They include studies on phase diagrams at zero temperature
\cite{lorenz,oyamatsu,ohy,sumiyoshi,gentaro1,gentaro2,williams}
and at finite temperatures \cite{lassaut}.
These earlier works have confirmed that, for various nuclear models,
the nuclear shape changes in the way of Eq.\ (\ref{sequence})
as predicted by Refs.\ \cite{hashimoto,rpw}.

In these earlier works, however,
a liquid drop model or the Thomas-Fermi approximation
is used with an assumption on the nuclear shape
(except for Ref.\ \cite{williams}).
Thus the phase diagram at subnuclear densities
and the existence of the pasta phases should be examined
without assuming the nuclear shape.
It is also noted that
at temperatures of several MeV, which are relevant to the collapsing cores,
effects of thermal fluctuations
on the nucleon distribution are significant.
However, these thermal fluctuations cannot be described properly by
mean-field theories such as the Thomas-Fermi approximation
used in the previous work \cite{lassaut}.

In contrast to the equilibrium properties, dynamical or non-equilibrium aspects
of the pasta phases had not been studied until recently 
except for some limited cases \cite{formation,review}.
Thus it had been unclear even whether or not the pasta phases can be formed 
dynamically within the time scale of the cooling of neutron stars
nor whether or not the formation of the pasta phases and the transitions
between them can be realized under non-equilibrium conditions
in collapsing stellar cores.
Especially, the latter problem is highly non-trivial because
these processes are accompanied by drastic changes of nuclear shape.

To solve the above problems, molecular dynamic approaches
for nucleon many-body systems (a comprehensive overview is given by
Ref.\ \cite{feldmeier-romp}) are suitable.
They treat the motion of the nucleonic degrees of freedom
and can describe thermal fluctuations and many-body correlations
beyond the mean-field level.
Molecular dynamics for nucleons were originally developed
for studying nuclear reactions, especially the dynamics of
heavy ion collisions.
Nowadays there are various versions
including the fermionic molecular dynamics (FMD) \cite{fmd,feldmeier},
the antisymmetrized molecular dynamics (AMD) \cite{amd} and
the quantum molecular dynamics (QMD) \cite{aichelin-review,aichelin}, etc.,
which differ in the form of the trial wave functions.
They are now widely used in studies on nuclear structure as well as
nuclear reactions.

Among these models, QMD is the most practical for studying the pasta phases,
which allows us to treat large systems consisting of several nuclei.
The typical length scale $r_{\rm c}$ of half of the inter-structure
is $r_{\rm c} \sim 10$ fm and the density region of interest is
around half of the normal nuclear density $\rho_{0} = 0.165 {\rm\ fm}^{-3}$.
The total nucleon number $N$ necessary to reproduce $n$ structures
in the simulation box is $N \sim \rho_{0} (2r_{\rm c}n)^{3}$ (for slabs).
It is thus desirable to use $\sim 10^4$ nucleons
in order to reduce boundary effects.
Such large systems are difficult to be handled by FMD and AMD,
whose calculation costs increase as $\sim N^4$,
but are tractable for QMD, whose calculation costs increase as $\sim N^2$
like ordinary classical molecular dynamics calculations
(we will see the reason of this difference later).

\begin{figure}[t]
\begin{center}
\rotatebox{0}{
\resizebox{6cm}{!}
{\includegraphics{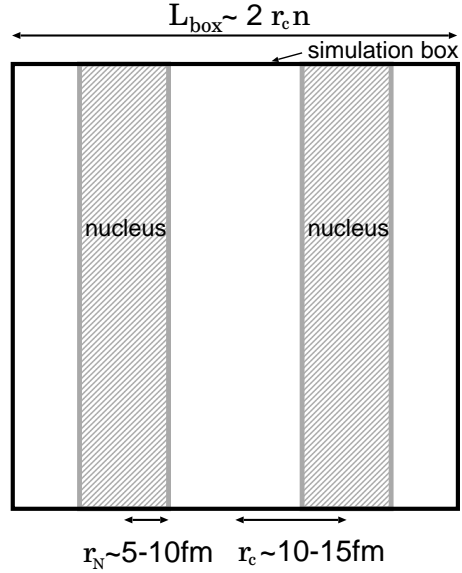}}}
\caption{\label{length scale}
  Typical length scales in the simulated system.
  The grey regions show ``pasta'' nuclei
  with the typical values of the half width $r_{\rm N} \sim$ 5--10 fm,
  and of the separation $r_{\rm c} \sim$ 10--15 fm.
  The typical value of
  the average nucleon density in the simulation box is about half of
  the normal nuclear density $\rho_{0}=0.165$ fm$^{-3}$.
  }
\end{center}
\end{figure}

It is also noted that we mainly focus on the nuclear structure
from mesoscopic to macroscopic scales (see Section \ref{sect_soft}),
where the exchange effect would be less important
\footnote{This can be seen by comparing the typical values of
the exchange energy for these larger scales
and of the energy difference
between two successive phases with non-spherical nuclei.
See, e.g., Ref.\ \cite{qmd2} for detail.}.
Therefore, it is expected that QMD is a reasonable approximation 
for studying the pasta phases even though it is less elaborate
on the treatment of the exchange effect than FMD and AMD.
Especially, at finite temperatures of several MeV,
validity of QMD is ensured because the shell structure 
(see Refs.\ \cite{chamel,shell n,shell p} for the shell effects in the pasta phases)
caused from the exchange effect is washed out by thermal fluctuations
above $T\sim 3$ MeV.

\subsection{Framework of QMD\label{subsect_frame}}

A general startpoint of constructing molecular dynamic approaches for nucleons
is the time dependent variational principle:
\begin{eqnarray}
  0 &=& \delta \int_{t_{1}}^{t_{2}} dt\
  \langle\Psi(t)| i\hbar \frac{d}{dt} - \hat{H} |\Psi(t)\rangle\ ,
\label{tdvp1}
\end{eqnarray}
where the time integral in the right hand side is the action.
If the variation $\delta$ of the state $\Psi(t)$ or $\Psi^{*}(t)$
is unrestricted in the whole Hilbert space
except for the fixed end points at time $t_1$ and $t_2$,
Eq.\ (\ref{tdvp1}) gives the usual time dependent Schr\"odinger equation.
Let us now restrict the space to be considered
by taking some trial many-body state $|\Phi \{Q(t)\}\rangle$,
which contains a set of dynamical variables (complex in general)
$Q(t)=\{ q_{1}(t),\ q_{2}(t),\ \cdots \}$.

In QMD, we assume each single-particle state $\phi_i$ of nucleon $i$
is a Gaussian wave packet and take the centers of positions
and momenta of the wave packets,
${\bf R}_{i}(t)$ and ${\bf P}_{i}(t)$, as a set of the dynamical variables
$Q(t)=\{ {\bf R}_{i}(t),\ {\bf P}_{i}(t);\ i=1,\cdots, N \}$;
\begin{equation}
  \phi_{i}({\bf r}) = \langle{\bf r}|\phi_{i}\rangle
     = \frac{1}{(2\pi \sigma^2)^{3/4}}
     \exp{\left[ -\frac{({\bf r}-{\bf R}_{i})^{2}}{4 \sigma^2}
       +\frac{i}{\hbar}\ {\bf r}\cdot{\bf P}_{i} \right]}\ ,\label{packet}
\end{equation}
where $\sigma$ is a parameter related to the extension of the wave packet
in the coordinate space.
Thus, $\sigma$ is set $\sim O(1)$ fm for nucleons
in nuclear matter at around the normal nuclear density.
The total $N$-nucleon wave function $|\Phi\{Q(t)\}\rangle$ is constructed as
a direct product of single-nucleon states defined by Eq.\ (\ref{packet})
\footnote{In FMD and AMD, $|\Phi\{Q(t)\}\rangle$ is 
a Slater determinant of single-nucleon states
instead of a direct product of them. Therefore, computational costs
for calculating the expectation value of the Hamiltonian in FMD and AMD are 
larger than those in QMD by a factor of $N^2$.}
\begin{equation}
  |\Phi\rangle = |\phi_{1}\rangle \otimes |\phi_{2}\rangle
     \otimes \cdots \otimes |\phi_{N}\rangle\ .\label{trial function}
\end{equation}

Employing the time-dependent variational principle
with the above trial state $|\Phi\rangle$ yields
the following equations of motion (QMD EOM):
\begin{equation}
\begin{array}{ccr}
  \dot{\bf R}_{i} &=&
     {\displaystyle \frac{\partial {\cal H}}{\partial {\bf P}_{i}}}\ ,\\ \\
  \dot{\bf P}_{i} &=&
     {\displaystyle -\frac{\partial {\cal H}}{\partial {\bf R}_{i}}}\ ,
\end{array}\label{qmdeom}
\end{equation}
where ${\cal H}(\{Q(t)\})$ is
the expectation value of the Hamiltonian $\hat{H}$:
\begin{equation}
  {\cal H} \equiv \langle\Phi| \hat{H} |\Phi\rangle\ .
  \label{hamilton func}
\end{equation}
We note that the above semiclassical equations of motion,
which incorporate the effect of the uncertainty principle,
have the same form as the classical Hamilton's equations.

The trial wave function of QMD given by Eq.\ (\ref{trial function})
does not possess antisymmetric properties.
Thus it is necessary to incorporate effects of the Pauli exclusion principle
in some way to reproduce fermionic nature:
one of the solutions is introducing the Pauli potential $V_{\rm Pauli}$.
The Pauli potential, a two-body repulsive potential, which depends on
the relative momentum as well as the relative coordinate distance,
mimics the Pauli principle by reproducing the exchange repulsive force
phenomenologically, i.e., by suppressing the phase space density
of identical particles.

There are several forms of the Pauli potential;
the following Gaussian form is one example:
\begin{equation}
  V_{\rm Pauli}=C_{\rm P} \left( \frac{\hbar}{q_{0} p_{0}} \right)^{3}\
  \exp{\left[ -\frac{({\bf R}_i-{\bf R}_j)^2}{2q_0^2} 
    -\frac{({\bf P}_i-{\bf P}_j)^2}{2p_0^2} \right]}\ ,\label{gaussian pauli}
\end{equation}
where $C_{\rm P},\ q_{0}$ and $p_{0}$ are parameters.
The parameters $q_{0}$ and $p_{0}$ determine
the range of the phase space distance
where the exchange repulsion acts.
They are thus set so that their product equals to
the volume element of the phase space: $q_{0} \cdot p_{0} \sim h$.

However, we have to mention that the genuine effect of the Pauli principle
caused by antisymmetrization is a $N$-body correlation,
not a two-body effect.
Therefore, the Pauli potential which try to imitate the Pauli principle
by two-body interaction cannot exactly describe quantum correlations
and quantum statistical properties.
Using this framework, we thus cannot obtain correct results for quantities
sensitive to the quantum fluctuations in the many-body system,
such as the specific heat at low temperatures \cite{ohnishi}.

\subsection{Model Hamiltonian\label{subsect_model}}

In our studies on the pasta phases, we have used a QMD model Hamiltonian
developed by Maruyama {\it et al.} \cite{maruyama}.
This model Hamiltonian consists of four parts:
\begin{equation}
  {\cal H} =
  K+V_{\rm Pauli}+V_{\rm nucl}+V_{\rm Coulomb}\ ,
  \label{hamiltonian}
\end{equation}
where $K$ is the kinetic energy;
$V_{\rm Pauli}$ is the Pauli potential of the Gaussian form given
by Eq.\ (\ref{gaussian pauli});
$V_{\rm nucl}$ is the nuclear interaction which contains
the Skyrme potential (an effective zero-range nuclear interaction),
symmetry energy,
and momentum dependent potential; $V_{\rm Coulomb}$ is the Coulomb energy
between protons.

The parameters in the Pauli potential are determined by fitting the kinetic energy
of the free Fermi gas at zero temperature.
The above model Hamiltonian reproduces 
the properties of nuclear matter around the normal nuclear density
(i.e., the saturation density, binding energy,
symmetry energy, and incompressibility),
and of finite nuclei in the ground state,
especially of heavier ones relevant to our interest \cite{kido,maruyama}.
Other properties of this Hamiltonian, which are important for the pasta phases 
(e.g., the surface tension) are investigated in Ref.\ \cite{qmd2}.
It is also confirmed that a QMD Hamiltonian close to this model
provides a good description of nuclear reactions including the low energy
region (several MeV per nucleon) \cite{niita}.

\section{Simulations and Results\label{sect_result}}

Using the framework of QMD, we have solved the following two major questions
\cite{qmd_transition,qmd1,qmd2,qmd_hot}.
1) Whether or not the pasta phases are formed by cooling down hot
uniform nuclear matter in a finite time scale smaller than
that of the neutron star cooling?
2) Whether or not transitions between the pasta phases can occur
by compression of matter during the collapse of a star?
In the present section, we would like to review our these works.

In our simulations,
we treated the system which consists of neutrons, protons, and electrons
in a cubic box with periodic boundary condition.
The system is not magnetically polarized,
i.e., it contains equal numbers of protons (and neutrons) with spin up and
spin down.
Relativistic degenerate electrons which ensure charge neutrality
can be regarded as a uniform background because electron screening is
negligibly small at relevant densities around $\rho_0$ 
\cite{maruyama_screen,review,screening}.
Consequently, one must take account of the long-range nature of the Coulomb
interaction (see also a discussion in Ref.\ \cite{tours}).
We calculate the Coulomb interaction by the Ewald summation procedure \cite{allen,frenkel}
taking account of the Gaussian charge distribution of the proton wave packets
(see Appendix A in Ref.\ \cite{qmd2}).

\subsection{Realization of the Pasta Phases from Hot Uniform Matter
\label{subsect_realization}}

In Refs.\ \cite{qmd1,qmd2}, we showed that the pasta phases are produced
from hot uniform nuclear matter and we studied phase diagrams
at zero temperature.
In these works, we first prepared a uniform hot nucleon gas
at a temperature $T \sim 20$ MeV as an initial condition
equilibrated for $\sim 500 - 2000$ fm$/c$ in advance.
To realize the ground state of matter,
we then cooled it down slowly until the temperature got $\sim 0.1$ MeV
or less for $O(10^{3}-10^{4})$ fm$/c$,
keeping the nucleon number density constant.
In the cooling process, we mainly used the frictional relaxation method
(equivalent to the steepest descent method), which is given by
the physically grounded QMD equations of motion [Eq.\ (\ref{qmdeom})]
plus small friction terms:
\begin{equation}
\begin{array}{ccr}
  \dot{\bf R}_{i} &=&
     {\displaystyle \frac{\partial {\cal H}}{\partial {\bf P}_{i}}
       - \xi_{R} \frac{\partial {\cal H}}{\partial {\bf R}_{i}}}\ ,\\ \\
  \dot{\bf P}_{i} &=&
     {\displaystyle -\frac{\partial {\cal H}}{\partial {\bf R}_{i}}
       - \xi_{P} \frac{\partial {\cal H}}{\partial {\bf P}_{i}}}\ .\label{qmdeom fric}
\end{array}
\end{equation}
Here the friction coefficients $\xi_{R}$ and $\xi_{P}$ are positive definite,
which determine the relaxation time scale.

\begin{figure}[t]
\begin{center}
\resizebox{12cm}{!}{\includegraphics{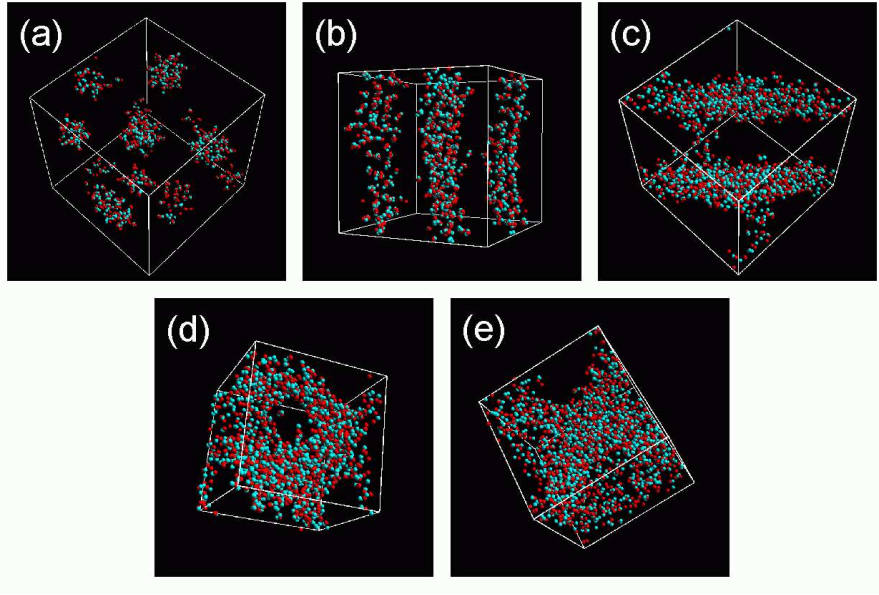}}
\caption{\label{fig pasta sym}(Color)\quad
  Nucleon distributions of the pasta phases in cold matter at $x=0.5$;
  (a) sphere phase, $0.1 \rho_{0}$ ($L_{\rm box}=43.65\ {\rm fm}$, $N=1372$);
  (b) cylinder phase, $0.225 \rho_{0}$ ($L_{\rm box}=38.07\ {\rm fm}$, $N=2048$);
  (c) slab phase, $0.4 \rho_{0}$ ($L_{\rm box}=31.42\ {\rm fm}$, $N=2048$);
  (d) cylindrical hole phase, $0.5 \rho_{0}$ ($L_{\rm box}=29.17\ {\rm fm}$, $N=2048$) and
  (e) spherical hole phase, $0.6 \rho_{0}$ ($L_{\rm box}=27.45\ {\rm fm}$, $N=2048$),
  where $L_{\rm box}$ is the box size and $N$ is the total number of nucleons.
  The whole simulation box is shown in this figure.
  The red particles represent protons and the green ones neutrons.
  Taken from Ref.\ \cite{qmd2}.
  }
\end{center}
\end{figure}

\begin{figure}[t]
\begin{center}
\resizebox{12cm}{!}{\includegraphics{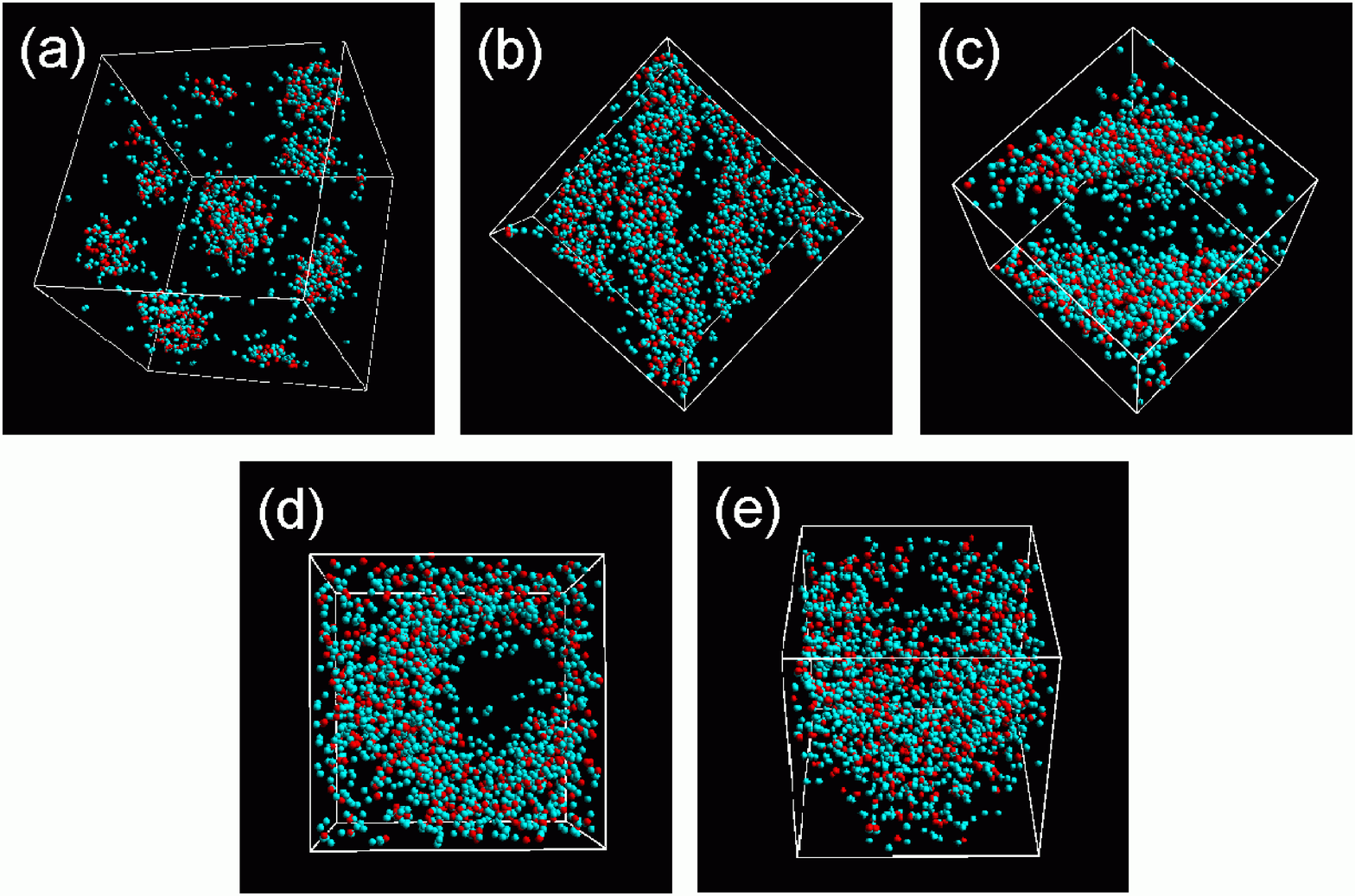}}
\caption{\label{fig pasta x0.3}(Color)\quad
  The same as Fig.\ \ref{fig pasta sym} at $x=0.3$;
  (a) sphere phase, $0.1 \rho_{0}$ ($L_{\rm box}=49.88\ {\rm fm}$, $N=2048$);
  (b) cylinder phase, $0.18 \rho_{0}$ ($L_{\rm box}=41.01\ {\rm fm}$, $N=2048$);
  (c) slab phase, $0.35 \rho_{0}$ ($L_{\rm box}=32.85\ {\rm fm}$, $N=2048$);
  (d) cylindrical hole phase, $0.5 \rho_{0}$ ($L_{\rm box}=29.17\ {\rm fm}$, $N=2048$) and
  (e) spherical hole phase, $0.55 \rho_{0}$ ($L_{\rm box}=28.26\ {\rm fm}$, $N=2048$).
  The red particles represent protons and the green ones neutrons.
  Taken from Ref.\ \cite{qmd2}.
  }
\end{center}
\end{figure}

The resulting typical nucleon distributions of cold matter 
at subnuclear densities are shown
in Figs.\ \ref{fig pasta sym} and \ref{fig pasta x0.3}
for proton fraction of matter $x=0.5$ and 0.3, respectively.
We see from these figures that
the phases with rod-like and slab-like nuclei,
cylindrical and spherical bubbles,
in addition to the phase with spherical nuclei are reproduced.
The above simulations show that the pasta phases 
can be formed dynamically from hot uniform matter
within a time scale $\tau \sim O(10^{3}-10^{4})$ fm$/c$.

\subsection{Equilibrium Phase Diagrams\label{subsect_phasediagram}}

\begin{figure}[htbp]
\begin{center}
\resizebox{13cm}{!}
{\includegraphics{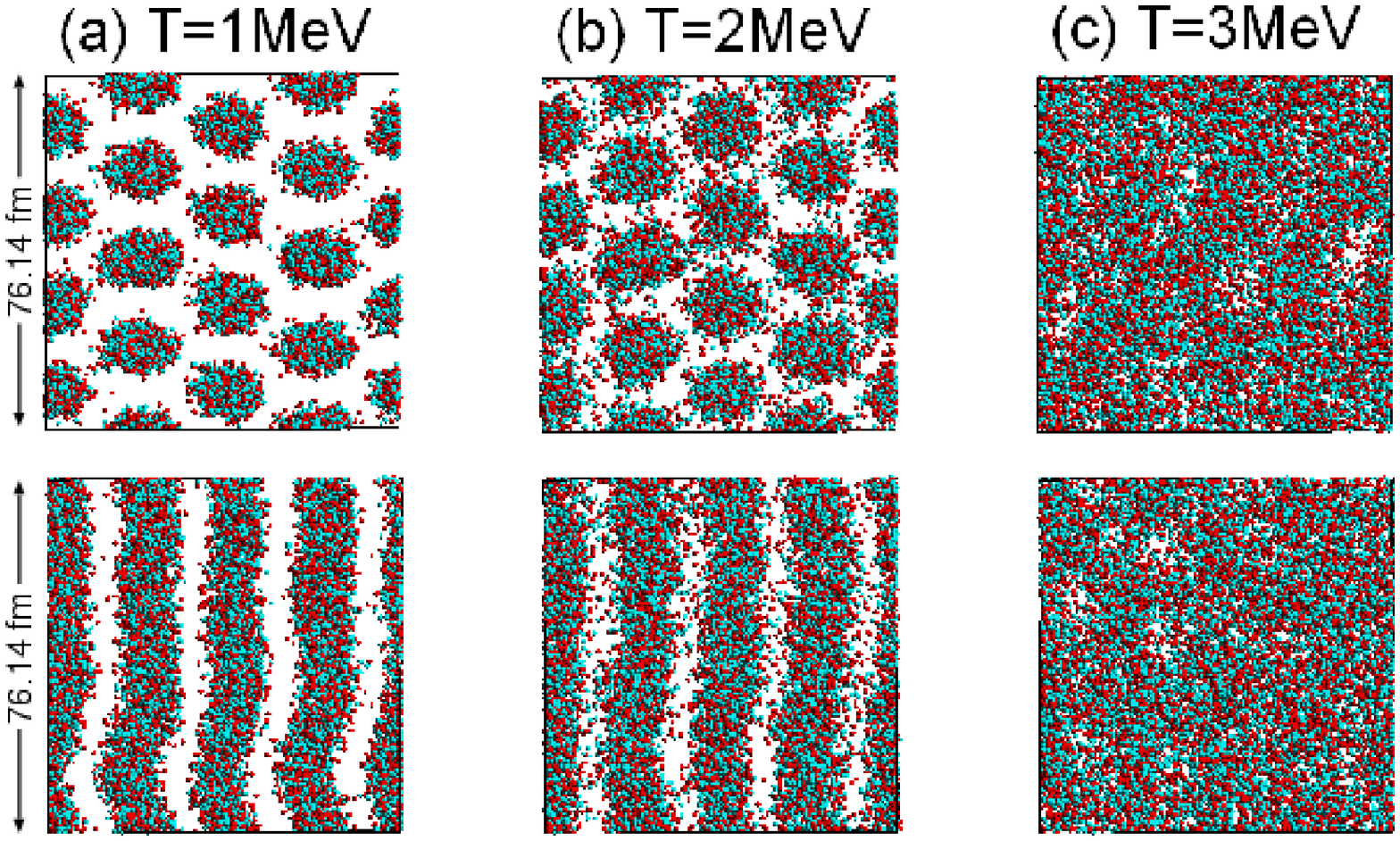}}
\caption{\label{snap 0.225rho x0.5 16000}(Color)\quad
  Nucleon distributions for $x=0.5$, $\rho=0.225\rho_{0}$
  at temperatures of 1, 2 and 3 MeV.
  The total number of nucleons $N=16384$ 
  and the box size $L_{\rm box}=76.14$ fm.
  The upper panels show top views along the axis of 
  the cylindrical nuclei at $T=0$, and the lower ones show side views.
  Protons are represented by the red particles, and
  neutrons by the green ones.
  Taken from Ref.\ \cite{qmd_hot}.
  }
\end{center}
%
\begin{center}
\resizebox{13cm}{!}
{\includegraphics{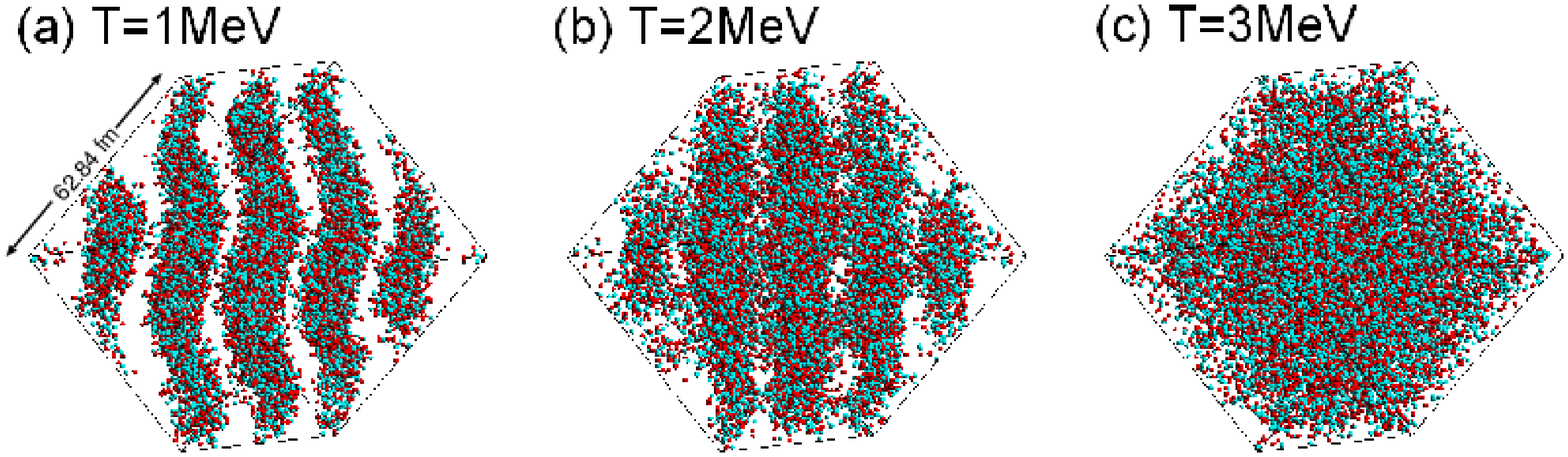}}
\caption{\label{snap 0.4rho x0.5 16000}(Color)\quad
  Nucleon distributions for $x=0.5$, $\rho=0.4\rho_{0}$
  at temperatures of 1, 2 and 3 MeV. $N=16384$ and $L_{\rm box}=62.84$ fm.
   These figures are shown in the direction parallel
  to the plane of the slab-like nuclei at $T=0$.
  Protons are represented by the red particles, and
  neutrons by the green ones.
  Taken from Ref.\ \cite{qmd_hot}.
  }
\end{center}
\end{figure}

As has been mentioned in Section \ref{sect_md_nucleon},
QMD provides a good description of thermal fluctuations
and is suitable for investigating the pasta phases at finite temperatures.
We show snapshots of nucleon distributions at $T=1, 2$ and 3 MeV
for two densities in Figs.\ \ref{snap 0.225rho x0.5 16000}
and \ref{snap 0.4rho x0.5 16000}.  These densities correspond to
the phases with rod-like nuclei and with slab-like nuclei at $T=0$.

From these figures, we can see the following qualitative features
(these features are common for $x=0.5$ and 0.3):
at $T\simeq 1.5-2$ MeV
(but snapshots for $T\simeq 1.5$ MeV are not shown),
the number of evaporated nucleons becomes significant; 
at $T\gtrsim 3$ MeV, 
nuclei almost melt and the spatial distribution of nucleons
are almost smoothed out.

When we try to classify the nuclear structure systematically,
the integral mean curvature and the Euler characteristic
(a brief explanation is provided in Appendix \ref{sect_euler})
are useful.  Here we introduce their normalized quantities:
the area-averaged mean curvature
$\langle H \rangle$,
and the Euler characteristic density $\chi / V$
($V$ is the volume of the whole space).
Using a combination of these two quantities calculated for nuclear surface
\footnote{Nuclear surface generally corresponds to an isodensity surface
for the threshold density $\rho_{\rm th}\simeq 0.5\rho_0$ 
in our simulations.}, 
each pasta phase can be represented uniquely, i.e.,
\begin{equation}
\mbox{phase with}\left\{
\begin{array}{ccc}
  \mbox{spherical nuclei}    &:& \langle H \rangle > 0,\ \chi/V > 0,\\
  \mbox{cylindrical nuclei}  &:& \langle H \rangle > 0,\ \chi/V = 0,\\
  \mbox{slab-like nuclei}     &:& \langle H \rangle = 0,\ \chi/V = 0,\\
  \mbox{cylindrical bubbles} &:& \langle H \rangle < 0,\ \chi/V = 0,\\
  \mbox{spherical holes}     &:& \langle H \rangle < 0,\ \chi/V > 0.\\
\end{array}
\right.
\end{equation} 

We note that the value of $\chi/V$ for the ideal pasta phases
is zero except for the phase with spherical nuclei or spherical bubbles
with positive $\chi/V$; 
negative $\chi/V$ is not obtained for the pasta phases.

\begin{figure}[htbp]
\begin{center}
\rotatebox{270}{
\resizebox{6.5cm}{!}
{\includegraphics{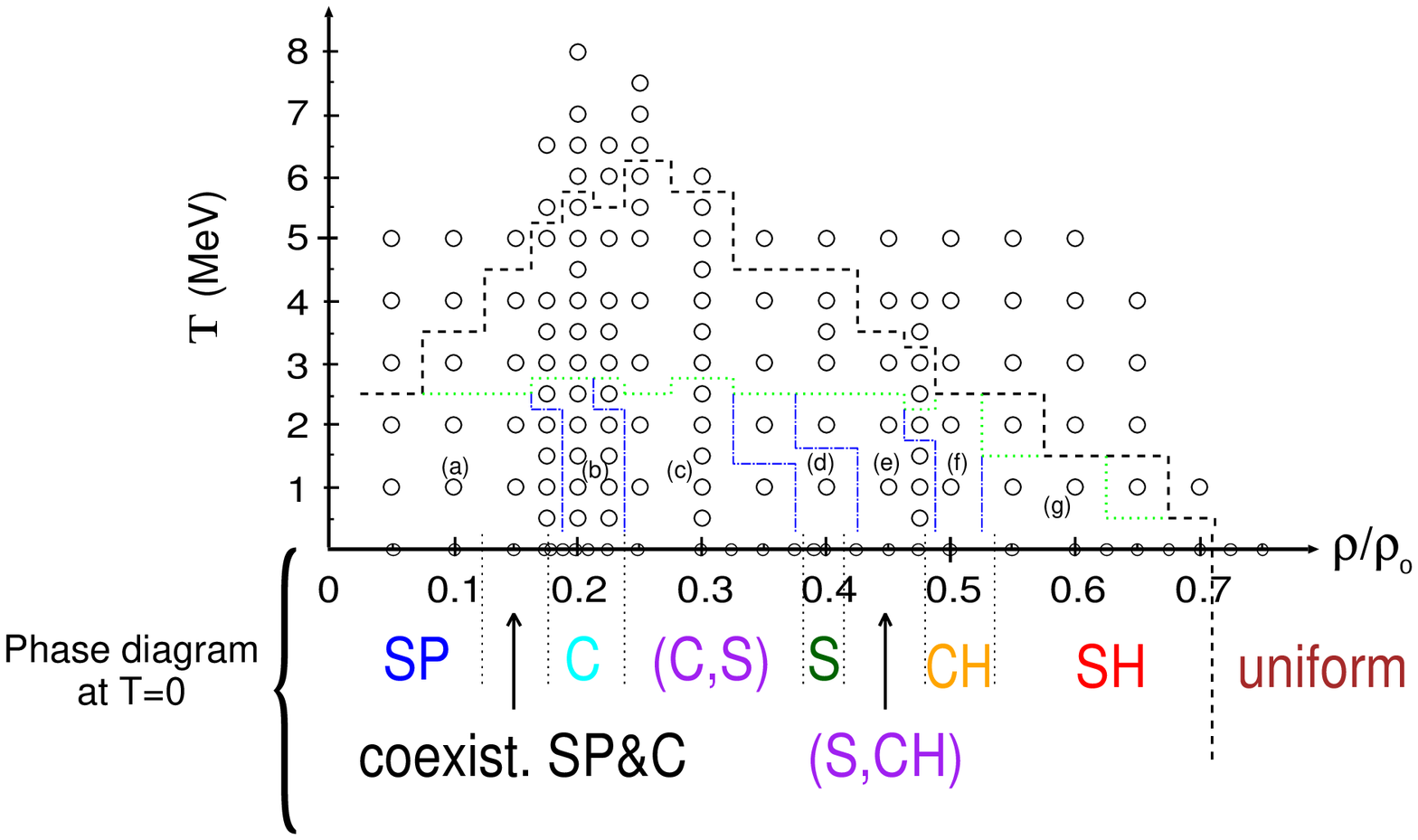}}}
\caption{\label{phase diagram x0.5}(Color)\quad
  Phase diagram of matter at $x=0.5$ plotted in the $\rho$ - $T$ plane.
  The dashed and the dotted lines on the diagram
  show the phase separation line and
  the limit below which the nuclear surface can be identified, respectively.
  The dash-dotted lines are the phase boundaries between
  the different nuclear shapes.
  The symbols SP, C, S, CH, SH, U stand for nuclear shapes,
  i.e., sphere, cylinder, slab, cylindrical hole,
  spherical hole and uniform, respectively.
  The parentheses (A,B) denote an intermediate phase between A and B-phases
  with a multiply connected structure characterized by 
  negative $\chi/V$.
  The regions (a)-(g) correspond to the nuclear shapes characterized by
  $\langle H \rangle$ and $\chi/V$ as follows:
  (a) $\langle H \rangle > 0,\ \chi/V > 0$;
  (b) $\langle H \rangle > 0,\ \chi/V = 0$;
  (c) $\langle H \rangle > 0,\ \chi/V < 0$;
  (d) $\langle H \rangle = 0,\ \chi/V = 0$;
  (e) $\langle H \rangle < 0,\ \chi/V < 0$;
  (f) $\langle H \rangle < 0,\ \chi/V = 0$;
  (g) $\langle H \rangle < 0,\ \chi/V > 0$.
  Simulations have been carried out at points denoted by circles.
  Adapted from Ref.\ \cite{qmd_hot}
  }
\end{center}
\end{figure}

\begin{figure}[htbp]
\begin{center}
\rotatebox{270}{
\resizebox{6.5cm}{!}
{\includegraphics{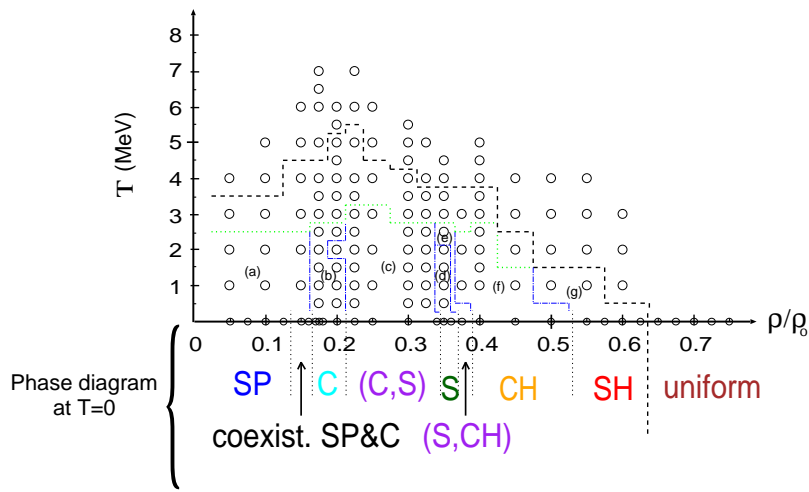}}}
\caption{\label{phase diagram x0.3}(Color)\quad
  Phase diagram of matter at $x=0.3$ plotted in the $\rho$ - $T$ plane.
  Adapted from Ref.\ \cite{qmd_hot}.
  }
\end{center}
\end{figure}

The phase diagrams obtained for $x=0.5$ and 0.3 are plotted in 
Figs.\ \ref{phase diagram x0.5} and \ref{phase diagram x0.3}, respectively.
Qualitative features of these phase diagrams are the same,
but the phase separating-region surrounded by a dashed line
is smaller for $x=0.3$ than that for $x=0.5$.
As shown above,
nuclear surface can be identified
typically at $T \lesssim 3$ MeV (see the dotted lines)
in the density range of interest.
Thus the regions between the dotted line and the dashed line
correspond to some non-uniform phase,
which is however difficult to classify
into a specific phase because the nuclear surface is hard to identify.

In the region below the dotted lines at $T\lesssim 3$ MeV,
where we can identify the nuclear surface, we have obtained 
the pasta phases with spherical nuclei [region (a)], 
rod-like nuclei [region (b)], slab-like nuclei [region (d)],
cylindrical holes [region (f)] and spherical holes [region (g)].
In addition to these pasta phases,
structures with negative $\chi/V$ have been also obtained
in the regions (c) and (e); matter consists of 
highly connected nuclear and bubble regions (i.e., sponge-like structure) with
branching rod-like nuclei, perforated slabs and branching bubbles, etc.
A detailed discussion on the phase diagrams is given in Ref.\ \cite{qmd_hot}.
Effects of the finite box size on the density regions of the phases
with negative $\chi/V$ should be examined in the future.
If these density regions do not decrease significantly even for larger systems,
the phases with multiply connected structures would be likely.

\begin{figure}[t]
\begin{center}
\rotatebox{270}{
\resizebox{3.6cm}{!}
{\includegraphics{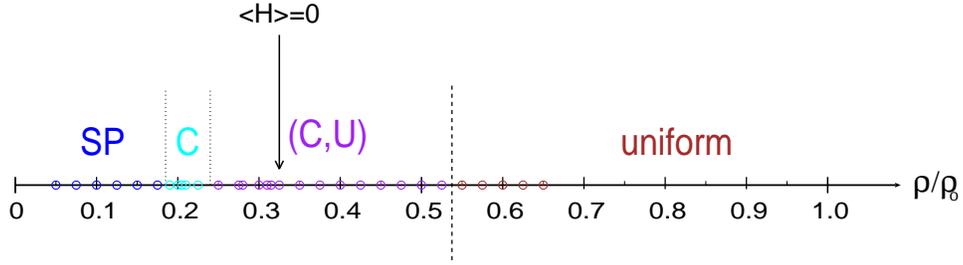}}}
\caption{\label{phase diagram x0.1}(Color)\quad
  Phase diagram of cold matter at $x=0.1$.
  The density at which the area-averaged mean curvature of nuclear surface
  is zero is denoted by $\langle H\rangle =0$
  ($\langle H\rangle >0$ at lower densities than this point
  and $\langle H\rangle <0$ at higher densities).
  However, slab phase is not observed in our results even at this density.
  Simulations have been carried out at densities denoted by small circles.
  Adapted from Ref.\ \cite{qmd2}.
}
\end{center}
\end{figure}

We have also investigated cold neutron-rich matter at $x=0.1$ 
as a more realistic condition for neutron star matter \cite{qmd2}.
The resulting zero-temperature phase diagram is shown
in Fig.\ \ref{phase diagram x0.1}.
The pasta phase with rod-like nuclei is obtained around $0.2 \rho_0$.
A striking feature is that a wide density region
from $\sim 0.25 \rho_{0}$ to $\sim 0.525 \rho_{0}$
is occupied by phases with negative $\chi/V$.
The structure of matter changes rather continuously from
branching rod-like nuclei connected to each other
[obtained in the lower density region
of the intermediate phase denoted by (C,U)]
to branching bubbles connected to each other
[higher density region of the intermediate phase (C,U)].
However, in the present neutron-rich case, pasta phase with slab-like nuclei
cannot be obtained as far as we have investigated.
A further study with a longer relaxation time scale
is necessary to determine whether or not the phase with slab-like nuclei
is really prohibited in such neutron-rich matter in the present model.

\subsection{Structural Transitions between the Pasta Phases
\label{subsect_transition}}

In Ref.\ \cite{qmd_transition}, we approached the second question
asked at the beginning of the present section.
We perform QMD simulations of the compression of dense matter
and have succeeded in simulating the transitions between 
rod-like and slab-like nuclei and between slab-like 
nuclei and cylindrical bubbles.

The initial conditions of the simulations are samples of the columnar phase
($\rho=0.225 \rho_0$) and of the laminar phase ($\rho=0.4 \rho_0$)
with 16384 nucleons at $x=0.5$ and $T\simeq1$ MeV.
These are obtained in our previous work \cite{qmd_hot},
which are presented in Section \ref{subsect_phasediagram}.
We then adiabatically compressed the above samples by increasing the density
at the average rate of $\simeq$1.3$\times 10^{-5} \rho_0/$(fm$/c$)
for the initial condition of the columnar phase and
$\simeq$7.1$\times 10^{-6} \rho_0/$(fm$/c$) for that of the laminar one.
According to the typical value of the density difference
between each pasta phase, $\sim 0.1\rho_0$ 
(see Fig.\ \ref{phase diagram x0.5}),
we increased the density to the value corresponding to the next pasta phase
taking the order of $10^4$ fm$/c$. This time scale is much longer than
the typical time scale of the deformation and structural transition of nuclei 
(e.g., that of nuclear fission is $\sim 1000$ fm$/c$).
Thus the simulated compression process is adiabatic 
with respect to the change of nuclear structure, so that the dynamics
of the structural transition of nuclei observed in the simulations is
physically meaningful, and is essentially independent of 
the compression rate, etc.
Finally, we relaxed the compressed sample
at $\rho=$0.405 $\rho_0$ for the former case and at 0.490 $\rho_0$ 
for the latter one.
These final densities are those
of the phase with slab-like nuclei and cylindrical bubbles, respectively,
in the equilibrium phase diagram at $T \simeq 1$ MeV 
(see Fig.\ \ref{phase diagram x0.5}).

\begin{figure}[t]
\begin{center}
\resizebox{13cm}{!}
{\includegraphics{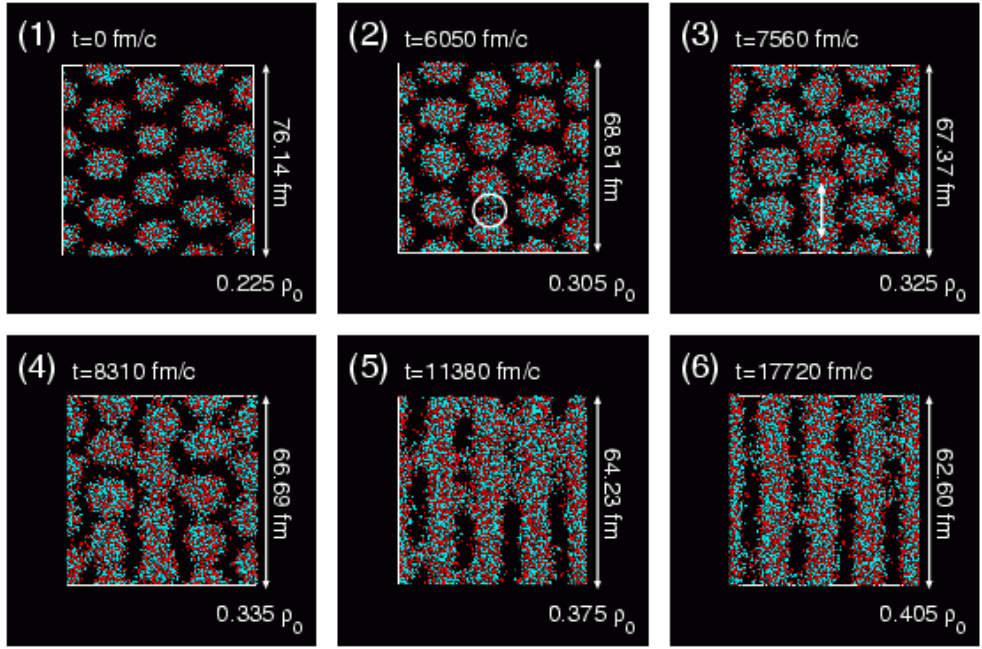}}
\caption{\label{fig rod slab}(Color)\quad 
Snapshots of the transition process from
the phase with rod-like nuclei to the phase with slab-like nuclei
(the whole simulation box is shown).
The red particles show protons and the green ones neutrons.
After neighboring nuclei touch as shown by the circle in 
Fig.\ \ref{fig rod slab}-(2),
the ``compound nucleus'' elongates along the arrow in 
Fig.\ \ref{fig rod slab}-(3).
The box size is rescaled to be equal in this figure.
Adapted from Ref.\ \cite{qmd_transition}.
}
\end{center}
\end{figure}

\begin{figure}[t]
\begin{center}
\resizebox{13cm}{!}
{\includegraphics{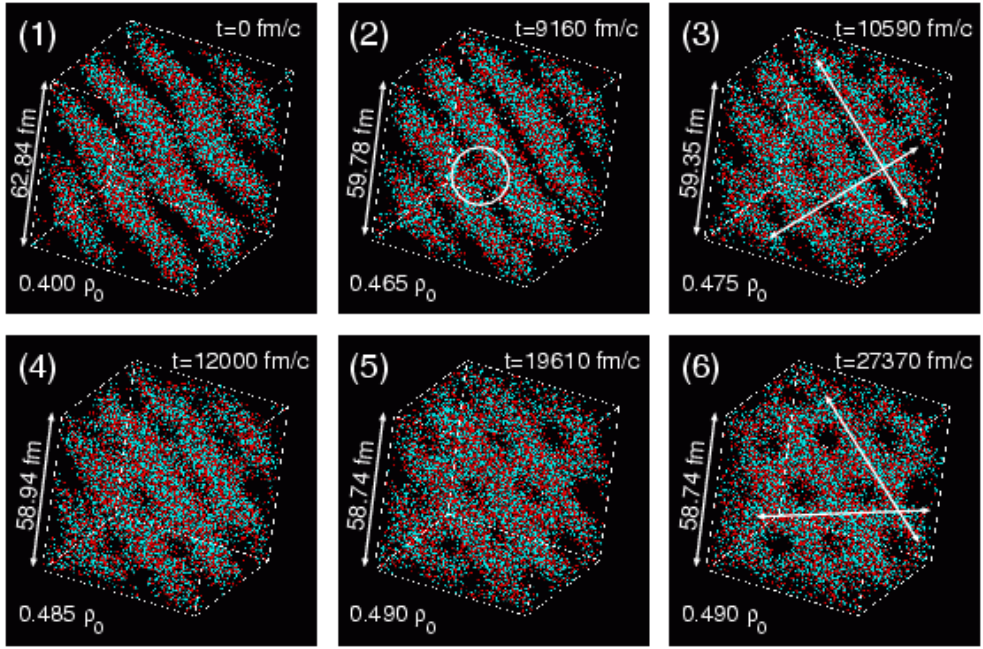}}
\caption{\label{fig slab cylindhole}(Color)\quad 
The same as Fig.\ 1 for the transition from
the phase with slab-like nuclei to the phase with cylindrical holes
(the box size is not rescaled in this figure).
After the slab-like nuclei begin to touch [see the circle in 
Fig.\ \ref{fig slab cylindhole}-(2)],
the bridges first crosses them almost orthogonally as shown by the arrows
in Fig.\ \ref{fig slab cylindhole}-(3). 
Then the cylindrical holes are formed and they
relax into a triangular lattice, as shown by the arrows in 
Fig.\ \ref{fig slab cylindhole}-(6).
Adapted from Ref.\ \cite{qmd_transition}.
}
\end{center}
\end{figure}

The resulting time evolution of the nucleon distribution
is shown in Figs.\ \ref{fig rod slab} and \ref{fig slab cylindhole}.
As can be seen from Fig.\ \ref{fig rod slab},
the phase with slab-like nuclei is finally formed 
[Fig.\ \ref{fig rod slab}-(6)]
from the phase with rod-like nuclei [Fig.\ \ref{fig rod slab}-(1)].
The temperature in the final state is $\simeq 1.35$ MeV.
We note that the transition is triggered by thermal fluctuation,
not by spontaneous deformation of the rod-like nuclei:
when the internuclear spacing becomes small enough 
and once some pair of neighboring rod-like nuclei 
touch due to thermal fluctuations,
they fuse [Figs.\ \ref{fig rod slab}-(2) and \ref{fig rod slab}-(3)].
Then such connected pairs of rod-like nuclei further touch and fuse with 
neighboring nuclei in the same lattice plane like a chain reaction
[Fig.\ \ref{fig rod slab}-(4)]; the time scale of the each fusion process
is of order $10^2$ fm$/c$, which is much smaller than 
that of the density change.

The transition from the phase with slab-like nuclei
to the phase with cylindrical holes is shown in 
Fig.\ \ref{fig slab cylindhole}.
When the internuclear spacing decreases enough,
neighboring slab-like nuclei touch due to the thermal fluctuation
as in the above case.
Once nuclei begin to touch [Fig.\ \ref{fig slab cylindhole}-(2)],
bridges between the slabs are formed at many places
on a time scale (of order $10^2$ fm$/c$)
much shorter than that of the compression.
After that the bridges cross the slabs
almost orthogonally for a while [Fig.\ \ref{fig slab cylindhole}-(3)].
Nucleons in the slabs continuously flow into the bridges,
which become wider and merge together to form cylindrical holes.
Afterwards, the connecting regions
consisting of the merged bridges move gradually,
and the cylindrical holes relax into a triangular lattice 
[Fig.\ \ref{fig slab cylindhole}-(6)].
The final temperature in this case is $\simeq 1.3$ MeV.

\begin{figure}[t]
\begin{center}
\rotatebox{270}{
\resizebox{!}{13cm}
{\includegraphics{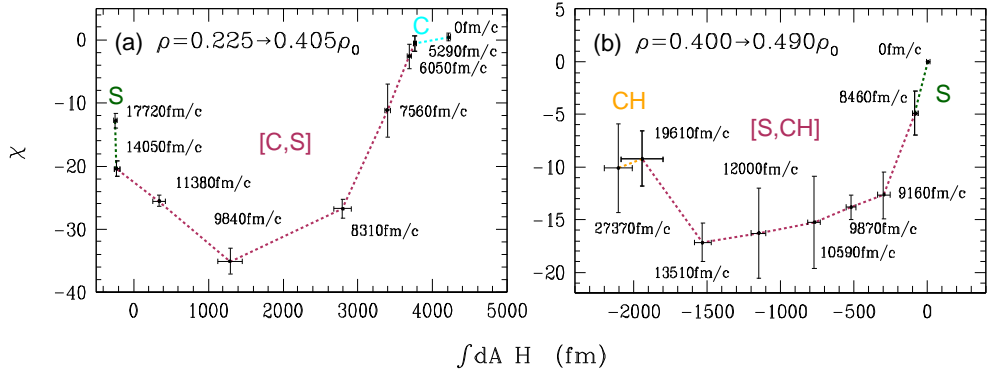}}}
\caption{\label{fig curv euler}(Color)\quad 
Time evolution of $\int_{\partial R} H dA$ and $\chi$
during the simulations.
The data points and the error bars show, respectively, the mean values
and the standard deviations in the range of the threshold density
(see Appendix \ref{sect_euler})
$\rho_{\rm th}=0.3$ -- $0.5 \rho_0$,
which includes typical values for the nuclear surface.
The panel (a) is for the transition from cylindrical (C)
to slab-like nuclei (S)
and the panel (b) for the transition from slab-like nuclei
to cylindrical holes (CH).
Transient states are shown as [C,S] and [S,CH] for each transition.
Adapted from Ref.\ \cite{qmd_transition}.
}
\end{center}
\end{figure}

Trajectories of the above processes on the plane of 
the integral mean curvature $\int_{\partial R} H dA$ 
and the Euler characteristic $\chi$ are plotted in Fig.\ \ref{fig curv euler}.
This figure shows that the above transitions proceed through a transient
state with ``sponge-like'' structure, which gives negative $\chi$.
As can be seen from Fig.\ \ref{fig curv euler}-(a) 
[Fig.\ \ref{fig curv euler}-(b)],
the value of the Euler characteristic starts to decrease from zero
when the rod-like [slab-like] nuclei touch. It continues to decrease
until all of the rod-like [slab-like] nuclei are connected to others
by small bridges at $t\simeq 9840$ fm$/c$ [$\simeq 12000$ fm$/c$].
Then the bridges merge to form slab-like nuclei [cylindrical holes]
and the value of the Euler characteristic increases towards zero.
Finally, the system relaxes into a layered lattice
of the slab-like nuclei [a triangular lattice of the cylindrical holes].
Thus the whole transition process can be divided into
the ``connecting stage'' and the ``relaxation stage''
before and after the moment at which the Euler characteristic is minimum; 
the former starts when the nuclei begin to touch and it
takes $\simeq 3000$ -- 4000 fm$/c$ and the latter takes
more than 8000 fm$/c$.

\subsection{Formation of the Pasta Phases
\label{subsect_formation}}

\begin{figure}[t]
\begin{center}
\resizebox{10cm}{!}
{\includegraphics{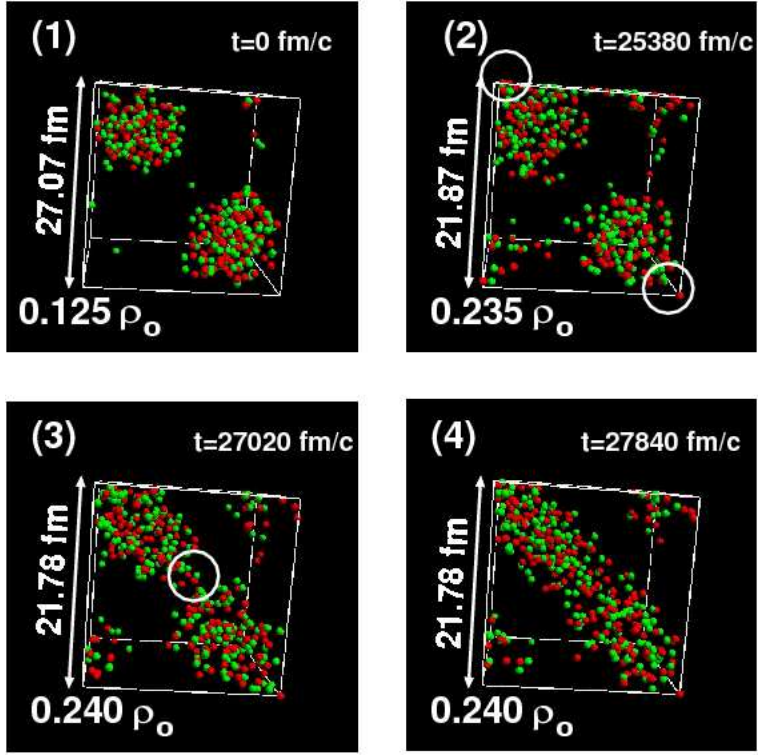}}
\caption{\label{fig sphere rod}(Color)\quad 
Snapshots of the transition process from
the bcc lattice of spherical nuclei to the triangular lattice of rod-like nuclei
(the whole simulation box is shown).
The red particles show protons and the green ones neutrons.
The box size is rescaled to be equal in this figure.
}
\end{center}
\end{figure}

In closing the present article, let us briefly show our recent results
of a study on the formation process of pasta nuclei
from spherical ones; i.e., a transition from the phase with spherical nuclei
to that with rod-like nuclei.
Time evolution of the nucleon distribution in the transition process
is shown in Fig.\ \ref{fig sphere rod}.
The initial condition of this simulation is 
a nearly perfect bcc unit cell with 409 nucleons (202 protons and
207 neutrons) at $T\simeq 1$ MeV.
We compressed the system in a similar way to that of the simulations
explained in Section \ref{subsect_transition}.
The average rate of the density change in the present case is
$\simeq$4.4$\times 10^{-6} \rho_0/$(fm$/c$).
Since the two nuclei start to touch [see the circles in 
Fig.\ \ref{fig sphere rod}-(2)], the transition process completes
within $\simeq 2500$ fm$/c$ and the rod-like nucleus is formed.
The final state [Fig.\ \ref{fig sphere rod}-(4)]
is a triangular lattice of the rod-like nuclei.

The present simulation has been performed using a small system;
effects of the finite system size in this simulation should be examined.
Detailed investigation of the transition using a larger system
will be presented in a future publication \cite{future}.

\section{Summary\label{sect_summary}}

In this article, we have explained the basic physics and astrophysics
of non-spherical nuclei, nuclear pasta, in compact objects.
In supernova cores and neutron stars, density can be very high;
when the density approaches that of atomic nuclei,
the Coulomb force starts to matter and to compete 
against the surface tension of nuclei.
Exotic nuclear shapes result from the competition 
between these two opposite effects.
The presence of the pasta phases would affect various astrophysical phenomena.
Especially, impacts on the neutrino opacity of collapsing stellar cores
should be examined in the near future.

We have discussed the similarity between the pasta phases
and soft condensed matter systems.
Pasta phases with rod-like nuclei and slab-like nuclei have similar
elastic properties to those of some phases of liquid crystals: 
columnar phase and smectic A phase, respectively.
With the help of this concept,
thermodynamic stability of the phase with slab-like nuclei 
and melting temperature
of the phase with rod-like nuclei have been discussed.

Nuclear systems can be regarded as complex fluids of nucleons
and their nature of a complex fluid sets limits 
on the validity of methods commonly used in previous studies of pasta phases.
However, a gap between the typical length scale of 
the semimacroscopic pasta phases and that of the nucleon is not large.
Due to this small gap, we can study pasta phases using
some microscopic approach: the quantum molecular dynamics 
(QMD) is one possibility.

We have explained the theoretical framework of molecular dynamics calculations for nucleons
including QMD and then given an overview of our previous work.
Our QMD simulations have shown that the pasta phases can be formed 
dynamically by cooling down hot uniform nuclear matter in a finite time scale
$\sim O(10^{3}-10^{4})$ fm$/c$, which is much shorter than the cooling time
of a neutron star.
We have also shown that transitions between the pasta phases can occur
by compression during the collapse of a star.
Our latest result strongly suggests that the pasta phase with rod-like nuclei
can be formed by compressing a bcc lattice of spherical nuclei;
this result will be examined in detail.

As we have shown,
nuclear pasta is an interesting system for a wide spectrum of researchers
not only to nuclear astrophysicists.
Physics of the pasta phases draws on nuclear physics, 
condensed matter physics, and astrophysics.
The pasta phases are important for understanding stellar collapse and
neutron star formation, which are long-standing mystery in the Universe.

\subsection*{Acknowledgements}
The authors are grateful to Kazuhiro Oyamatsu for giving us permission
to use Fig.\ \ref{pasta}.
They also thank Chris Pethick for helpful comments.
The research reported in this article grew out of collaborations with
Kei Iida, Toshiki Maruyama, Katsuhiko Sato, Kenji Yasuoka and
Toshikazu Ebisuzaki.
Further research currently in progress is performed using the RIKEN
Super Combined Cluster System with MDGRAPE-2.
This work was supported in part
by the Nishina Memorial Foundation,
by a JSPS Postdoctoral Fellowship for Research Abroad,
by a JSPS Research Fellowship for Young Scientists,
by the Ministry of
Education, Culture, Sports, Science and Technology
through Research Grant No. 14-7939,
and by RIKEN through Research Grant No. J130026.

\appendix

\section{The Integral Mean Curvature and the Euler Characteristic
\label{sect_euler}}

The integral mean curvature and the Euler characteristic
(see, e.g., Ref.\ \cite{minkowski} and references therein for details)
are powerful tools for extracting the morphological characteristics
of the structure of nuclear matter.
Suppose there is a set of regions $R$,
where the density is higher than a given threshold density $\rho_{\rm th}$. 
The integral mean curvature and the Euler characteristic
for the surface of this region $\partial R$
are defined as surface integrals of
the mean curvature $H = (\kappa_{1}+\kappa_{2})/2$ and
the Gaussian curvature $G = \kappa_{1} \kappa_{2}$, respectively;
i.e., $\int_{\partial R} H dA$ and
$\chi \equiv \int_{\partial R} G dA/2\pi$,
where $\kappa_{1}$ and $\kappa_{2}$ are the principal curvatures and
$dA$ is the area element of the surface of $R$.
The Euler characteristic $\chi$ depends only on the topology of $R$
and is expressed as
\begin{equation}
  \chi = \mbox{(number of isolated regions)}
  - \mbox{(number of tunnels)} + \mbox{(number of cavities)}\ .
  \label{euler}
\end{equation}
Their normalized quantities, the area-averaged mean curvature
$\langle H \rangle$ and the Euler characteristic density
are also useful. They are defined as
$\langle H \rangle \equiv A^{-1}\int H dA$,
and $\chi / V$, respectively,
where $V$ is the volume of the whole space.

According to Eq.\ (\ref{euler}), sphere and spherical holes give $\chi>0$;
infinitely long rods, infinitely extending slabs and
infinitely long cylindrical holes give $\chi=0$.
Multiply connected structures such as ``sponge'' give negative $\chi$.


\label{lastpage-01}
\end{document}